# 'Setting Fire to the Last Forest': Project West Ford and the Mobilization of the Astronomical Community 1958–1965

John C. Barentine
Dark Sky Consulting, USA

## Abstract

The dawning of the Space Age marked the start of an ongoing relationship between the professional astronomical community and both state and non-state actors that launch and operate spacecraft in near-Earth orbital space. While the Cold War heated up in the late 1950s, military uses of outer space quickly came into conflict with the priorities of astronomers then building ever-bigger ground-based telescopes and envisioning the first generation of space telescopes. As the threat of global thermonuclear war loomed, the United States carried out Project West Ford, which tested an 'artificial ionosphere' for microwave radio propagation by placing several hundred million tiny copper dipoles into a belt orbiting the Earth. While the test was ultimately successful, it ignited a firestorm of concern and criticism among astronomers and ultimately influenced the framing of the United Nations Outer Space Treaty. Here we examine the history of Project West Ford as it prompted astronomers to react, comparing it with the ongoing problem of the potential impact of large satellite constellations on astronomical research.

## Keywords

# Introduction

For nearly seven decades, astronomy and astrophysics have existed in a world where outer space has been continuously accessible to humans. While this enabled powerful space telescopes that have greatly increased our understanding of the cosmos, other uses of space have come to constitute a threat to the continued access to cosmic light from on and near the Earth's surface. Most recently, the rapid commercial development of low-Earth orbit (LEO) space by both public and private entities poses a new threat to astronomy.[1] The deployment of large satellite constellations in LEO, and the space debris they can generate, present a challenge to observational astronomy unprecedented in its history.[2] Astronomers reacted to the arrival of large constellations, quickly finding limits to their ability to shape policy and practice relating in part to their failure to anticipate the problem.

One may well ask whether there are any historical parallels that inform the ongoing relations between astronomers, the governments of launching states, and the companies that design, launch and operate satellites. Project West Ford (PWF) is a particularly instructive case study. PWF was a United States military operation of the early 1960s that attempted to create an 'artificial ionosphere' above the Earth to ensure stable microwave radio communications over large distances in the event that natural radio propagation were disrupted due to the detonation of one or more nuclear weapons in or above the atmosphere. While the U.S. military endeavored to consult with scientists in the planning and execution of PWF, astronomers expressed concerns about potential deleterious impacts to their optical, infrared and radio-wavelength observations on Earth as well as the potential for long-term persistence of the artificial ionosphere that could hinder planned future space telescopes.

Most importantly for understanding the present situation with large satellite constellations, the astronomy community decried the lack of consultation on the part of the military planners. Although the single successful test of PWF did not yield any discernible impairment of ground-based astronomical observatories, it raised questions about the safe and equitable use of outer space for various purposes during the burgeoning 'Space Race' between the United States and the Soviet Union in the 1960s. This helped frame the provisions of the Outer Space Treaty (1967), which became the cornerstone of the modern international space policy framework.

In this paper, we trace the history and development of PWF from inception to conclusion, with particular emphasis on the participation of the U.S. scientific community and the protests of non-U.S. astronomers whose work stood to be affected by American military decisions. We then explain how the affair influenced the development of the Outer Space Treaty, the international legal framework that presently governs uses of space. Lastly, we identify lessons from PWF that may inform the ongoing relationship between astronomers and the operators of large satellite constellations in the interest of protecting the integrity of the global ground-based astronomy enterprise.

# Origins of Project West Ford

Global radio communications began around the turn of the 20th century and were routine by the start of the Second World War. Before the satellite era, long-range telecommunications depended in part on natural radio propagation in the Earth's upper atmosphere. Space weather events could substantially disrupt the ionosphere, affecting high-frequency (HF) radio propagation and causing radio blackouts. On the other hand, some instances such as the severe geomagnetic storm of 14 May 1921 apparently enhanced propagation even as it damaged telegraph equipment.[3] Even in fair conditions, radio propagation was a function of many variables that reduced its reliability. During the interwar period, scientists and engineers sought ways to make radio communications more robust against these natural disruptions.[4] This was especially important in the pre-satellite era, when long range HF connections were essential for military purposes.

## *The Cold War context (1945–1957)*

The end of the Second World War saw a rapid realignment of world powers that resulted in a new geopolitical order pitting the Soviet Union and its client states in Eastern Europe against the United States and its allies in Western Europe. The destructive power of nuclear weapons and emerging belief in so-called mutually assured destruction in the next large-scale armed conflict yielded an effective stalemate between East and West lasting throughout the following half-century.

Building on the rapid industrialization of both the U.S. and U.S.S.R. during the war years, both sides looked to science as the key to achieving cultural, military and economic superiority. Billed in America as the "endless frontier",[5] supporting fundamental scientific research through public investment was seen as both a catalyst of innovation as well as a matter of national security. As a result, scientists took a place on the front line of the Cold War, often finding that they had to "go along to get along" with military priorities in this era. Few held positions of sufficient influence to push back against this regime in meaningful ways. This led to what Stewart Leslie calls an emerging "military-industrial-academic complex" at American universities in the 1950s.[6]

Harnessing the power of the atom for purposes both peaceful and bellicose defined the period during and immediately after the Second World War. German scientists discovered nuclear fission in 1938; its theoretical basis led to the idea that fission could power weapons yielding destruction on a scale unprecedented in the history of warfare. Concerned about the possibility that Nazi Germany would achieve this capability first, in a series of actions from 1940–42 the U.S. committed to the Manhattan Project, a program of research and development intended to effectively counter the Nazi nuclear threat. The promise and peril of nuclear weapons was realized in the Trinity test on 16 July 1945, followed within weeks by the atomic bombings of Japan and its surrender shortly thereafter. Already plans were underway to deliver nuclear weapons over very long distances. In 1944, German rockets became the first objects made by humans to reach the threshold of space, paving the way for long-range and intercontinental ballistic missiles that would soon be tipped first with conventional atomic bombs and later by thermonuclear devices.

As rocket technology able to lift heavier payloads to new heights progressed, the nuclear weapons they could carry grew ever more powerful. Concerns emerged about the potential for electromagnetic pulses resulting from atmospheric nuclear detonations and the temporary disruption of conditions in the ionosphere critical to global radio propagation.[7] At the time, long-range communications were possible only via undersea cable or HF radio transmissions, which were considered more militarily secure. A deliberate act to effectively destroy long-distance

"skywave" propagation might leave an adversary unable to communicate with military assets such as submarines in the field. Given that early Cold War ideas about a full thermonuclear exchange between the military superpowers of the day identified the decisive advantage of a "first strike" posture, the potential information blackout that could result from a nuclear airburst was elevated to a high-level concern on both sides.

## *Military developments (1957–1959)*

The Soviet launch of *Sputnik 1* on 4 October 1957 surprised the American public.[8] But it was particularly significant given the powerful launch vehicle required. In a speech at Chatham House, London, on 17 December 1957, British Member of Parliament Denis Healey summarized the West's anxiety at the development: "From the military point of view, the sputnik means that Russia has the capacity to produce a missile which is capable of carrying a thermonuclear warhead a distance of some five thousand miles in something like twenty minutes, and of guiding that missile with sufficient accuracy to destroy the Capitol building in Washington."[9]

Having been defeated by the U.S.S.R. in the race to reach outer space first, the U.S. government interpreted the advent of space travel as an existential threat if not countered and redoubled its efforts to achieve a presence in orbit. The U.S. Navy's attempt to launch the Vanguard 1A satellite in early December 1957 ended in failure, but the test paved the way for the successful launch of *Explorer 1* the following month. It revealed the first hints of the Van Allen radiation belts around Earth, confirmed during the mission of its successor *Explorer 3*, suggesting that the space environment near the Earth would make for challenging operating conditions for later spacecraft.[10]

## *Project Echo (1960-1964)*

Around the same time, in late 1957, American scientists and engineers began studying the idea of using Earth-orbiting satellites for long-distance communications at microwave frequencies.The need to establish reliable and robust radio communications over long distances no matter the vagaries of space weather was underscored by an intense solar storm on 11 February 1958. Occurring near the peak of Solar Cycle 19 — the most intense solar maximum in two centuries — the storm caused an extensive radio blackout over North America and demonstrated that even undersea cable communications could be disturbed by major geomagnetic events.[11] These incidents had the potential to disable or even destroy satellites.[12]

The launch of *Sputnik 1* also accelerated the pace of experimentation with space-based communications. In July 1958, the U.S. Air Force convened a conference on communications satellites. Bell Telephone Laboratories engineer John R. Pierce presented a concept for passive satellite relay of radio signals from point to point on the Earth by way of an orbiting reflector.[13] Pierce developed the idea further with his Bell Labs colleague Rudolf Kompfner under the guise of probing the refraction of radio waves through the atmosphere using large balloons deployed in low-Earth orbit and presented a paper on the design at National Symposium on Extended Range and Space Communication on 6-7 October.[14] Bell Labs backed the effort on the presumption that it would encourage development of communications relays in space. The newly established National Aeronautics and Space Administration, along with the Jet Propulsion

Laboratory (JPL), agreed to partner with Bell Labs on 'Project Echo' and set a target date of September 1959 for the first launch.[15]

The Echo spacecraft were balloons made of aluminized Mylar that were packed in a deflated state for launch and inflated to their maximum diameter of 30 m upon reaching orbit. The balloons were fully passive reflectors of radio energy, transmitting only a weak telemetry signal using battery-powered FM or VHF beacons. Three individual balloons were launched separately between 1960 and 1964, successfully testing two-way radio communications between the JPL Goldstone facility in the Mojave Desert of California and Bell Telephone Laboratories facility in Holmdel, New Jersey.[16] Given their size and highly reflective surfaces at both radio and optical wavelengths, the balloons were conspicuous to the unaided eye over much of the Earth.

The engineering achievements included the first two-way telephone conversation carried by satellite on 15 August 1960, and the first satellite relay of a live television transmission in April 1962. Given this success, the U.S. proposed Project Rebound, which would have involved the launch of as many as 90 Echo-type balloons to create a global passive communications network.[17] It was intended as a stopgap measure in 1962–63 until NASA could launch satellites to geostationary orbit (GEO); however, before the program could be developed sufficient launch capability was attained and NASA's *Syncom 2* became the first satellite to reach GEO.

## *Atmospheric nuclear tests (1958)*

As the solar storm of February 1958 was underway, planning was well along for Operation Hardtack I, a U.S. military program that among other objectives aimed to study the prospect of countering the Soviet nuclear attack threat by disabling or destroying incoming ballistic missiles by means of detonating nuclear weapons at high altitude. It also intended to characterize the electromagnetic waves emanating from a nuclear explosion, which were expected to have damaging effects on electronic devices. One of the last tests in the series, code-named "Teak", was conducted on 1 August 1958. A Redstone missile lifted the payload to an altitude of 76.8 km above the North Pacific Ocean where it detonated directly over Johnston Atoll with an energy yield approximately 156 times that of the first U.S. Army nuclear test in 1945. Large quantities of fission products from the explosion were injected into the ionosphere, which disrupted normal HF radio propagation. This caused a radio blackout over the Pacific Ocean lasting several hours, which halted commercial trans-Pacific air transportation.[18] It was accompanied by a brilliant display of light from an induced aurora.

This was shortly followed by Operation Argus, which intended to test the entrapment of relativistic electrons produced by neutron and beta decay of fission fragments in the Earth's magnetic field by detonating nuclear weapons in space.[19] Military strategists believed it possible to repel the effects of Soviet nuclear missiles by detonating a few nuclear weapons at high altitudes over the South Pacific Ocean. The confined cloud of electrons would destroy the electronics of Soviet warheads as they flew through the region. By ionizing a denser region of the atmosphere below the ionosphere, engineers hoped that the effect might also blind Soviet radars to weapons launched in a U.S. counterstrike. The tests proved successful in generating the temporary low-altitude radiation belts, but the effect dissipated too quickly to convey any meaningful strategic advantage.[20] It also elicited protests from the scientific community because the tests were conducted under conditions of military secrecy, which meant that many scientists were entirely unaware of the significant electron enhancement in the upper atmosphere in the weeks following the tests.[21]

## Planning and engineering

### Project Barnstable (1958)

The prospect of future nuclear warfare in space and the development of missiles capable of lifting heavier payloads to higher altitudes pushed the U.S. to invest more in research on the upper atmosphere and the space environment in order to optimize space weapons. Fearing a loss of long-range radio propagation due to nuclear blasts, in spring 1958 the U.S. Army Signal Corps appealed to the Massachusetts Institute of Technology (MIT) to organize an effort to study the issue with the goal of issuing recommendations on how to make future military communications more robust. MIT convened a group of scientists and engineers who met weekly in a schoolhouse on Cape Cod during summer 1958. Their effort took its name from the nearby town of Barnstable, Massachusetts.[22]

The United States Space Science Board (SSB) of the National Academies of Science and a subcommittee of the President's Science Advisory Committee (PSAC) nominated members of the committee. Those individuals were scientists and science policy experts predisposed to favor the project despite what negative effects it might have on the U.S. scientific enterprise.[23] There is a strong historical parallel in the twenty years preceding the onset of the Second World War in which the U.S. military effectively controlled stratospheric ballooning in order to gain a competitive advantage; Tanya Levin notes that "scientists who wished to use the balloon for research had to learn to accommodate the missions and objectives of the military."[24] Similarly, newly qualified space scientists of the late 1950s and early 1960s would soon find their research interests pre-empted by military priorities.

During the meeting series, Harold F. Meyer (Ramo-Wooldridge Corporation) and Walter E. Morrow, Jr., (MIT's Lincoln Laboratory) proposed to improve the reliability of command communications in space by putting only passive components into orbit that would be easy to track from the ground. They identified the need for a continuous region of space that was reflective at microwave wavelengths — which the ionosphere is not — to serve as a natural relay of microwave signals. The carrier medium, in the words of Lincoln Laboratory scientist Irwin Shapiro, had to be "jam-proof and indestructible",[25] a requirement that did not characterize any artificial satellite of the era.

The medium would take the form of small filaments of copper wire, nicknamed 'needles', to be carried into orbit on an Air Force rocket and automatically dispensed. (Figure 1) The use of similar metallic "chaff" to either thwart enemy radar or enable tropospheric signal propagation over very remote areas was already well established by the early days of the Cold War.[26] The needles' dimensions were chosen so as to make them half-wave dipole reflectors at the resonant frequencies of incoming radiation between 7.75 and 8.35 GHz.

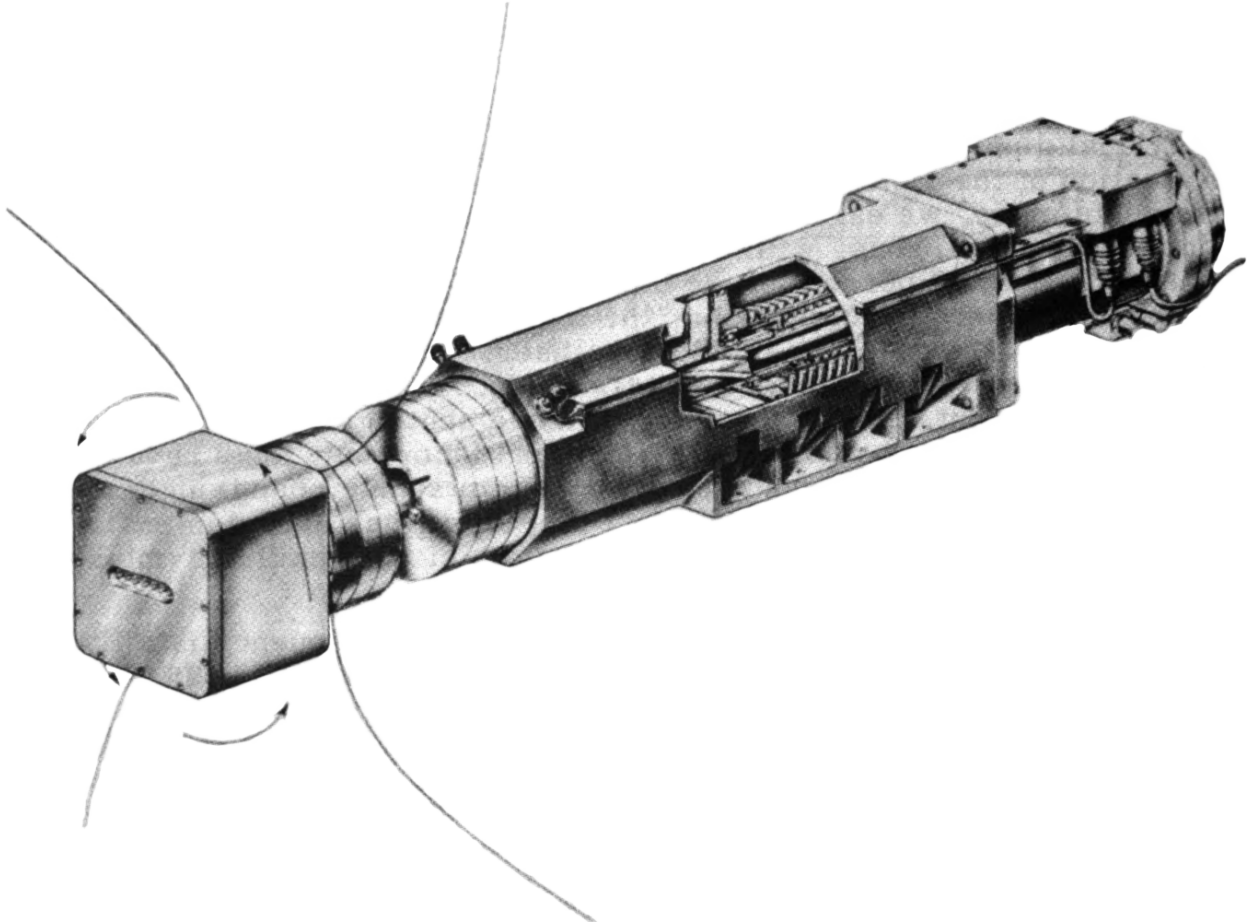

**Figure 1.** The Project West Ford dipole dispenser. Copper "needles" embedded in a block of solid naphthalene inside the device were pushed toward the rotating dispensing head at left.[27]

Within weeks of launch the dipoles would disperse along their orbit, forming a belt around the Earth. The belt would provide continuous radio connectivity yet be essentially invulnerable to Soviet attack. It would also ensure radio propagation at least at some frequencies in the event a catastrophic solar storm resulted in severe disruption of the ionosphere. The innate redundancy of the orbital dipole belt made it more resistant to enemy attack than active satellite relay systems, and its high (*c*. 3600 km) orbit situated the dipoles well above the highest recorded altitude of any nuclear detonation to date.[28]

The initially classified effort was initially known as "Project Needles". This designation in turn lent itself to the naming of MIT Haystack Observatory, used to test the efficacy of the dipole belt after its successful deployment: the difficulty of observing signals reflected from it was likened to 'finding a needle in a haystack'.[29] The mechanism proposed by Meyer and Morrow was received with skepticism by military officials, but the U.S. Air Force came to see value in the idea and offered to fund the project and carry the payload on one of its rockets. The Air Force named Morrow as the project's director; he quickly assembled a team of engineers that worked on methods for fabricating the dipoles and dispensing them on-orbit. New facilities with suitable radio antennas, transmitters and receivers were built near Boston, Massachusetts, and San Francisco, California, to test the system after deployment.[30]

Despite the design, according to Paul Rosen, later head of the ground equipment satellite group at Lincoln Laboratory, the dipole belt was "not very practical." Decades later he recalled that

> *its useful bandwidth was very, very small; the modulation system was very complex; and its cross section was tiny. While it would be a suitable vehicle for very important strategic communications like saying, "Go. Now is the time," it was not much good for anything else.*[31]

Almost immediately, concerns were raised about potential adverse impacts of the project on radio astronomy, optical astronomy and space travel. As a reflector of microwave radiation, the dipole belt would be opaque to astrophysical signals at those wavelengths, and would probably itself become a source of radiofrequency interference . It was feared that the dipoles would scatter light at optical wavelengths, affecting telescopes on the ground. And no one knew whether spacecraft venturing into the dipole belt would suffer collisions that could damage or destroy them. Astronomers also worried about what technology like the Needles would mean for space telescopes, which they believed would become available by the end of the 1960s. In 1961, Harvard University astronomy professor William Liller wrote of the project that "the chance of seeing for the first time extremely faint objects with satellite telescopes possessing resolving powers no longer limited by an atmosphere in a sky perfectly black except for the feeble zodiacal light at the ecliptic poles appears to be threatened."[32]

It was thought that the dipole belt might raise the radio continuum brightness of the sky by a temperature equivalent to about 5 K.[33] If scattering a directed beam, the brightness temperature of the sky in that direction might increase by as much as 3 K. Radio refraction effects, on the other hand, were expected to be small. Direct attenuation of optical light passing through the belt was thought to be very small, on the order of $10^{-10}$ times the initial intensity. Władysław Turski (Polish Academy of Sciences) estimated that the needles might create a diffuse glow through scattering of natural sources. For an assumed natural sky background luminance of 157 microcandela per square meter, the effect was expected to be about 1%.[34] The dipoles might be a problem for polarimetry if they became aligned with the Earth's magnetic field.[35]

There was concern from very early on about what unintended consequences the experiment might have on the space environment. Turski argued that

> *the successful implementation of this project would change in some, unfortunately not sufficiently precise, way the physical characteristics of the outer space surrounding the Earth. What's more — in the case of only a partial success of the experiment, i.e., if the process of dipole dissemination from the satellite or the orbit of the satellite itself deviated from the predictions of the authors of the project — the changes caused by the dipole cloud may be completely unpredictable. … Even a minor error in the design [of the orbit] can lead to colossal changes in the 'life' of the dipole belt. … It is [therefore] extremely difficult to give any answer to the question of what will be the consequences of the implementation of the West Ford project and what will be the 'life' of the dipole belt, which is an essential part of this project.*

Lincoln Laboratories staff reckoned that these potential hazards were likely to be negligible, "but the scientific community had to be convinced of the validity of this prediction."[36] Given the contemporaneous criticism from the scientific community on the military's handling of Operations Hardtack I and Argus, "it was clear from the beginning that the views and opinions of a variety of scientists should be sought."[37]

## *The Villard Committee (1960)*

In September 1959 Morrow traveled to New York with Carl Overhage, then the director of Lincoln Laboratory, and William H. Radford (MIT) to meet with Lloyd V. Berkner, chairman of the SSB. The Laboratory approached the SSB for an opinion about the potential effects of the Needles project on both scientific research and other activities in space.[38] Berkner suggested establishing an ad hoc committee of the SSB to study the question of whether the dipole belt would cause interference to astronomy. It would also consider the potential reactions of both the scientific community and the public, American and otherwise, to news about the project. The committee, formally appointed on 24 December, was to be chaired by Stanford University electrical engineering professor Oswald G. Villard, Jr. Astronomers participating as committee members included Gerald M. Clemence (U.S. Naval Observatory); Cornell H. Mayer (Naval Research Laboratory) and Fred L. Whipple (Smithsonian Astrophysical Observatory).

The committee met three times on 8 January, 12 February and 14 April 1960. By the third meeting, 33 scientists had participated in addition to Lincoln Laboratory staff. Those included John W. Findlay (National Radio Astronomy Observatory), who later observed that "the areas of scientific interest were mainly atmospheric science, astronomy (optical and radio), ionospheric physics, meteorites and radio propagation." At the April meeting, Findlay recalled,

> *it had become generally agreed that the first Needles experiment, using some 75 pounds of copper dipoles packed in naphthalene, was very unlikely to produce harmful effects or interference in any of the following respects:*
>
> *1) possible collision between individual needle dipoles and Project Mercury capsules,*
> *2) interference to operational or research radar installations,*
> *3) insolation of the earth,*
> *4) electron content of the ionosphere during re-entry of the dipoles, and*
> *5) contamination of the upper atmosphere by either the dipole or the binder material.*

Nevertheless,

> *there remained considerable doubts about the possible effects of orbiting dipole belts on optical and radio-astronomical measurements. These doubts were coupled with some uncertainty about the lifetime of a belt. However, so long as it was clear that only the one experimental belt was being considered, and so long as they were not asked to say that even the one experiment was either necessary or desirable, the optical and radio astronomers agreed that the belt described by Lincoln Laboratory would not seriously harm either optical or radio astronomy.*[39]

The committee sent a confidential report to the full Board on the 15th in which members noted that they were "deeply disturbed by the project ... because of what appears to be a serious threat to radio astronomy observations in the short-wave end of the radio spectrum."[40] It also feared that a successful test of PWF might well prompt both the U.S. and U.S.S.R. to each seek their own, more extensive dipole belts which could quickly crowd orbital space with material reflective at microwaves, effectively ending ground-based radio astronomy at such wavelengths.[41] In this anxiety the first hints of a future 'space environmentalism' emerged, complete with an anticipated — and potentially imminent — tragedy of the commons.[42] The report also envisioned a future in which multiple dipole belts were unnecessary, given that even one would be equally accessible to everyone, avoiding the worst outcome. Yet it ended with a recommendation that the Air Force terminate the project given its unclear effects on both optical and radio astronomy.

Over the following two months, committee members carried out and reported studies of how the needles were expected to behave in space and to interact with the space environment. Behind the scenes Berkner exerted pressure on individual members, due to politics more than science, to abandon their opposition to PWF given the power of the military authority backing it. Berkner understood the military's intent to proceed toward a test launch irrespective of any emergent opposition from the scientific community, believing that astronomers would wield more control over how the test unfolded if they maintained a sense of cooperation with the Air Force and Lincoln Laboratory.[43]

In its final 6 June report to the SSB, the Villard Committee abruptly changed its recommendation in favor of Project Needles. But it also urged transparency and openness to deflect negative public scrutiny; limiting the orbital lifetime of the dipole belt to "as few years as possible"; establishing protected frequency bands for radio astronomy; and searching for a more permanent solution — presumably, dedicated communication satellites. The SSB met near Los Angeles, California, in late June to review the report. Discussion among members of its Astronomy Committee, chaired on the occasion by Princeton University Observatory director Lyman Spitzer, was lively; Findlay remembered that they "went on through the day (in and around the swimming pool) and continued far into the night."

The SSB issued a statement on 30 June summarizing its view of the Needles project:

*1. It does not appear that the first exploratory test which has been proposed by Lincoln Laboratory will have an adverse effect on any branch of science. In the case of optical and radio astronomical observations, however, the Board is very much concerned with the harm which an operational system or systems might entail. This is true even though means may be found to give the dipoles a limited lifetime. We strongly recommend that any planning for an operational system or for large-scale future tests protect the interests of basic astronomical research and of science in general.*

*2. It is recommended that scientific and operational aspects of the initial experiment of Project Needles and its importance be declassified as soon as possible except for those relating to specific military uses. Release of information should be through recognized international channels, such as the international unions, COSPAR [Committee on Space Research], United Nations, and the scientific literature, and the positive aspects of the project should be demonstrated.*

*3. The Board recognizes the seriousness of the interference to radio astronomy which is likely to result not only from dipole belts, but also from active communication satellites. The Board favors the establishment of frequency bands for radio astronomy protected on a worldwide basis.*

*4. There should be established a committee of radio astronomers to work closely with the developers of Project Needles and to serve in an advisory capacity in connection with the question of interference to radio astronomy.*[44]

# Astronomers react

Looking back on the period during the run up to the PWF tests, Findlay later reflected that

> *it was by now evident that astronomers were concerned more for their future in the event of many dense belts being used than they were about the one initial Needles experiment. Some of the general argument was about tactics. There were fears expressed that once the experiment had proved successful, future applications might appear so desirable to defense communication systems planners that consideration of the scientific problems would take second place. Since future larger-scale developments might harm the science, would it be better to oppose at once the one Needles experiment? Or would it be better not to oppose the first apparently harmless experiment and to wait to see and measure its effects? This latter opinion prevailed, but only as a majority view.*[45]

The backdrop against which to evaluate the response is an increasingly fragmented U.S. astronomy landscape, a portion of which was "more dependent upon the military following the war" and that had to "adapt to, and compromise with, their military patrons who attached different objectives and goals—national security rather than scientific—to the particular effort."[46] Also, the origins of Project West Ford coincided with the 1957–58 International Geophysical Year (IGY), an international collaboration aimed to promote the development of space science, including upper-atmosphere and space research.[47] As the scientific community came together with the common goal of exploiting its increasing access to outer space as a means of better understanding our planet and its surroundings, ironically it faced potential constraints on those activities coming from other users of space.

## *The Findlay Committee (1960)*

Lincoln Laboratory engineers and scientists concluded that the potential harm to astronomy might be small if the project were carried out flawlessly and the orbital lifetime of the belt was near the approximately 7-year pre-launch estimate. However, many astronomers were unconvinced by this argument. Some were worried about unpredictable effects of consequences unforeseen in the design and execution, while others expressed concern over the effects that efforts like PWF might have in the future, given that they could not know what technologies would be available then. The Space Science Board could not definitively dismiss these fears, committing to study of the dipole belt and transparency in its communications with scientists.

Fulfilling the fourth element of its report, in July 1960 the SSB established a committee chaired by Findlay. The committee's membership was composed of Fred T. Haddock (University of Michigan); W. Albert Hiltner (Yerkes Observatory); William Liller; A. E. Lilley (Harvard College Observatory); and Alan R. Sandage (Mount Wilson and Palomar Observatories). George A. Derbyshire (National Acadamy of Sciences) represented the Space Science Board Secretariat until September 1961, after which the representative was astronomer and NAS staffer Edward R. "Ned" Dyer, Jr. The committee first met on 29 August at the Lincoln Laboratory. Shortly thereafter Project Needles was formally renamed "Project West Ford".[48] The Findlay Committee's first task was to learn as much as possible about the project and try to understand the predictions about the behavior of the dipole belt made by Lincoln Laboratory staff, an effort that Findlay described as "very exhaustive."

Almost immediately, internal disagreements emerged among the committee members with respect to the SSB recommendations. Findlay cited the example of Leo Goldberg, an astronomer and newly minted Harvard professor, who

> *was doubtful whether such a group could give the project sufficiently close study to reduce the fears of scientists. At his instigation, the optical astronomers were added, and he joined the group himself at several of its meetings.*[49]

Goldberg chaired the Astronomy Committee of the SSB and met with the Findlay Committee as a liaison of the Board.[50] He pursued the matter privately in correspondence with Berkner during the winter of 1960–61. Berkner continued to insist that positive cooperation between astronomers and the Air Force (through Lincoln Laboratory) was crucial to astronomers realizing any benefit from the deal. Goldberg played along on the condition that the orbital dipole belts would be subject to further scientific study before any launches occurred. He knew full well that such investigations required declassifying at least some information about Project Needles; otherwise, if the military resisted, astronomers would have a stronger moral basis (if not a scientific one) for their complaints.

The Committee's second main task was to open lines of communication with the scientific community and provide factual information about PWF to scientists. On 2 September, an International Council of Scientific Unions (ICSU) Inter-Union Commission was set up by the International Union of Radio Science (URSI) Executive Committee. British radio physicist John A. "Jar" Ratcliffe chaired the first meeting of the Inter-Union Commission at the thirteenth USRI General Assembly in London on the 5th. Morrow was sent by Lincoln Laboratory to attend. Ratcliffe suggested that coordination of activities with Commission V of URSI could be achieved if Findlay were to be co-opted to the committee as its secretary. He suggested that Findlay attend the second meeting scheduled for the 8th. Findlay obliged, presenting the report of Sub-Commission Ve on Frequency Allocation, established at the 1957 General Assembly of URSI, and outlining the work that had been done in obtaining legal protections for radio astronomy frequencies during 1958–59.[51] On the 9th, Morrow presented a paper titled "Orbital Scatter Communication" to the General Assembly. As a result, in September Commission V of URSI adopted a resolution "expressing grave concern as to possible damage to astronomical research by projects of this nature, and asked URSI, with the IAU [International Astronomical Union], to try to avoid such damage".[52]

By October 1960 the basic parameters of the PWF test were known and Morrow had communicated them to the scientific community. Shapiro and Harrison Jones, also of Lincoln Laboratory, wrote in *Science*:

> *Some 75 pounds of tiny hairlike copper dipoles are to be injected into a circular polar orbit about 3800 km above the earth's surface; the resonant frequency of the dipoles will be near 8000 Mcy/sec. Radio-frequency equipment located near San Francisco and Boston will be used to conduct communications tests and to study the physical and electrical characteristics of the belt by means of monostatic and bistatic radar experiments.*[53]

Despite research into "several possible deleterious effects which might result from such a dipole belt" showing such concerns to be "groundless", the decision was made to put the dipoles into an orbit that would decay quickly.[54]

The Findlay Committee met again with Morrow and his associates in January 1961. The Lincoln Laboratory staff provided an update on the project status and answered questions relating to certain parameters of the test.[55] In March, ahead of the first expected PWF test in May, optical astronomers began organizing a coordinated campaign to observe and measure sunlight scattered from the dipole belt. Liller, designated the Committee's liaison to the optical astronomy community, "distributed a circular letter to a large number of optical astronomers throughout the world (i) providing a technical description of the experiment, (ii) containing recommendations for an observational programme which might be undertaken, and (iii) inviting the co-operation of individual astronomers in obtaining observations and measurements."[56]

In response to the SSB recommendations, four articles about the expected impacts of PWF prepared by the Findlay Committee were published in *The Astronomical Journal* in April. To reach as wide an audience as possible, the SSB arranged to distribute 1400 reprints of the articles; 1100 were sent to individual members of the IAU, of which 800 were outside of the U.S. (Findlay, 1964) A note about West Ford was also printed in the preliminary announcement of the 108th Meeting of the American Astronomical Society (AAS) to be held in Nantucket, Massachusetts, from 18–21 June.

Liller summarized the effects of PWF on optical astronomy, recommending several investigations "which might be carried out before the launching date which would reduce some of the uncertainties" and suggesting the kinds of observations of the orbiting dipoles that should be made. Acknowledging that astronomers were then beginning to worry about the impacts on space-based astronomy platforms they anticipated might be launched into low-Earth orbit within the potential lifetime of the dipole belt, Liller commented that

> *the possibility of higher albedos occurring at shorter wavelengths, the existence of skies made far darker because of the absence of airglow and man-made lights, and possibly other factors, will make the influence of the dipoles on satellite astronomy more important than on earth-bound astronomy. The chance of seeing for the first time extremely faint objects with satellite telescopes possessing resolving powers no longer limited by an atmosphere in a sky perfectly black except for the feeble zodiacal light at the ecliptic poles appears to be threatened by Project West Ford.*[57]

He further admitted the potential for the dipole belt to affect the measurement of polarized light from astrophysical sources: "Any tendency of the 2-cm needles to align themselves parallel to each other, as may conceivably happen as they move along the earth's magnetic field, could very well introduce serious errors in similar observations made in the future."

Most importantly, Liller addressed head-on concerns that experiments like PWF were subject to government classification as state secrets:

> *It should be strongly emphasized that all the calculations described in this section are based on the values of needle density and rate of dispersion given by Morrow and MacLellan. It is not possible to check these important quantities, because the details of the orbit and of the dispensing mechanism remain classified. This is most unfortunate,*

*since the correctness of the predictions made here are of vital concern to all optical astronomers in the world.*[58]

## Protest among astronomers rises

By early summer the Findlay Committee felt confident that it understood the aims and methods of PWF well, and it was satisfied with the outreach effort undertaken to convey adequate factual information about the project to the astronomical community. But their sense of accomplishment did not last long, as it was "shaken somewhat by the reactions of some astronomers and by a further difficulty which arose in the plans for the first West Ford launch."[59] The national scientific academies of several countries had received expressions of concerns from their astronomers, and the academies of France and Belgium had issued statements indicating their anxiety about the possible effects of the dipole belt on science. Some astronomers were firmly opposed to dipole belts as a concept. Others accepted the predictions of negligible harm from the variety of test proposed in PWF but feared the deployment of future dipole belts that might be more dense.

In mid-June the AAS convened its 108th meeting, during which it considered the matter. Its Council was "gravely concerned at the harm that a future operational belt or belts might cause to astronomical observations" and on the 20th it adopted a resolution consisting of two statements:

*1. The American Astronomical Society strongly opposes any contamination of space that is detrimental to the conduct of basic scientific research.*

*2. The Society reaffirms its previous resolution in support of world-wide frequency allocations for radio astronomy as a matter of great and increased urgency.*[60]

Amateur astronomers were also alarmed by the prospect of a changing night sky resulting from PWF and similar programs. Writing for *The Review of Popular Astronomy* in the summer of 1961, the American amateur astronomer Alan McClure expressed dismay about Projects West Ford and Rebound, fearing for the future of ground-based astronomy.[61] Assuming that the PWF test belt "will probably be visible to our eyes under favorable conditions, [and] the later operational belts most certainly would be," he expected they would appear to observers on the ground as "two false narrow 'Milky Ways'." If the test were successful, McClure argued, both military and private commercial interests would want more of them. Those interests "cannot be relied upon to give adequate consideration to the preservation of the night sky for pure research and the contemplation of man." McClure concluded:

*In gaining space might man lose the beauty of the night sky? Astronomers, both amateur and professional, are increasingly concerned by new space projects being considered now. To gain the light of new knowledge, must we 'set fire to the last forest'?*[62]

In his article, McClure recounted his private correspondence with several astronomers in which he discussed his concerns with them. Aden Meinel (Kitt Peak National Observatory) wrote that "while it is hard to say that any one experiment is harmful, the trend is disturbing. ... One solution for astronomers is to plan on placing our future observatories in high orbits or on the rear of the Moon as a last resort." About the prospect of PWF brightening the night sky in the optical, an unnamed astronomer "said that he felt that 'an increase in the sky illumination of

as much as one percent for any period of time will hamper those working to the limit of their telescopes."'

Concerning "the need for preservation of the night sky", Henry C. Giclas (Lowell Observatory) said to McClure:

*Of course, I feel the greatest argument for control of what is put in the sky should be that astronomical observations have not only provided us with the fundamental laws of mechanics and order, but the most fundamental concepts of matter and energy. I do not believe we have begun to exhaust this physical laboratory which has opened such a new and radical concept to man's mind and imagination that he could never possibly have conceived or attempted experiments on earth that have brought us to this point of progress. To jeopardize this unique and fruitful source of original data with manmade contaminants is most short-sighted and presumptive on his part.*

McClure closed his article with a reflection by Frank Drake (National Radio Astronomy Observatory) on what efforts like PWF meant for the future of technology development:

*It is clear that space technology has reached a state where it has a potential to harm other fields of science and everyday life. A close watch must be kept to insure that dangerous undertakings are nipped before they occur. There will obviously be a need in the near future for an international forum in which all the ramifications of an experiment can be discussed and a consensus of international opinion obtained as to the pros and cons of an experiment, with a final opinion as to whether the experiment should be done at all. We can hope that mankind will face up to this responsibility and carry it out without allowing petty political considerations and propaganda opportunities to corrupt the purpose of the organization.*[63]

McClure pressed the matter with the U.S. government. On 22 October 1961 he sent a telegram to U.S. Vice President Lyndon B. Johnson expressing his concern about and disapproval of the Project, referring to the "shocking disregard for the world's astronomers" and the government's "seemingly arrogant refusal to cooperate." Johnson replied in a letter of 24 October:

*If this space experiment had not received such careful evaluation prior to approval having been given to its launch, there would be some possible explanation for your reaction. However, the facts are that the project was examined by many outstanding scientists, including astronomers. For example, the Space Science Board of the National Academy of Sciences concluded, after a thorough study, that the project would not have adverse effects upon astronomy or upon other space projects. Among their conclusions were that it would be relatively short-lived and that the quantity of 75 pounds was too small to have any significantly adverse effect. It was estimated that it would intercept only 'one billionth part of incoming light and one millionth part of the radio waves."*[64]

Shortly after McClure's article was published, in August 1961 the Council of the Royal Astronomical Society (RAS) expressed its concerns about the pending PWF test in a letter to Jerome B. Wiesner, chair of the U.S. President's Science Advisory Committee:

*The Council of the Royal Astronomical Society views with greatest concern the reported intention of the U.S. National Aeronautical and Space Administration to put into orbit a large number of reflecting dipoles. Such a step would begin to shut off one of the few*

*wavelengths available to radio astronomy to further our knowledge of distant regions of space. At the present stage of development of radio astronomy it is quite impossible to say that a particular wave-length could be dispensed with, and it would be scientifically a grave disaster if this project were permanent. It seems inconceivable to the Council that a drastic step so gravely affecting any science could be even contemplated without the fullest consultation with all astronomical and other bodies likely to be concerned.*[65]

Three days later, the White House issued a policy statement declaring that no further PWF tests would be conducted until the results of the initial launch had been thoroughly reviewed:

*The United States Government, in conducting the West Ford Project, will be guided as follows:*

*1. No further launches of orbiting dipoles will be planned until after the results of the West Ford experiment have been analyzed and evaluated. The findings and conclusions of foreign and domestic scientists (including the liaison committee of astronomers established by the Space Science Board of the National Academy of Sciences) should be carefully considered in such analysis and evaluation.*

*2. Any decision to place additional quantities of dipoles in orbit, subsequent to the West Ford experiment, will be contingent upon the results of the analysis and evaluation and the development of necessary safeguards against harmful interference with space activities or with any branch of science.*

*3. Optical and radio-astronomers throughout the world should be invited to cooperate in the West Ford experiment to ascertain the effects of the experimental belt in both the optical and radio parts of the spectrum. To assist in such cooperation, they should be given appropriate information on a timely basis. Scientific data derived from the experiment should be made available to the public as promptly as feasible after the launching.*[66]

Wiesner sent Berkner the policy statement on the 11th, noting in his cover letter that it had been "prepared by the National Aeronautics and Space Council and had been approved by the President."[67] On the same day, the Space Science Board communicated the results of its investigation to IAU Commission 50:

*1. The Project West Ford experiment will constitute no interference to optical or radio astronomy. As a matter of fact, the belt will be barely detectable, even by astronomers with advance information and upon the taking of special efforts for detection. It is true that a belt or belts could be erected which could cause serious interference to astronomical observations; however, the United States Government policy provides that no further launches of orbiting dipoles will be planned until the West Ford results have been analyzed and evaluated and further, will be contingent on the development of necessary safeguards.*

*2. The Board will continue its studies of this area of experimentation on behalf of the scientific community. In these studies it will depend on objective and quantitative assessments that constitute the foundation for scientific discussions, recommendations and decisions. These assessments can only be achieved through a carefully controlled, harmless test, and Project West Ford provides a clear opportunity for scientists of all*

*nations to cooperate in making observations to form the basis for an objective understanding of the behavior of an orbiting dipole belt, both in terms of its astronomical properties and of its communication capabilities.*

*3. The Board will continue to keep the scientific community everywhere informed and it invites the cooperation and assistance of scientists everywhere who have interest and specialized knowledge in this area. The Board acknowl- edges with gratitude the assistance of many scientists—both at home and abroad—who have already contributed to its studies.*[68]

### *The XIth IAU General Assembly (15-24 August, 1961)*

The IAU convened in Berkeley, California, in August for its triennial General Assembly. Findlay later wrote that the meeting "was the forum for some vigorous discussions of West Ford."[69] While many in attendance were opposed to PWF, "some were evidently hearing of it for the first time despite the care taken to inform all the members well in advance." In his opening address to the General Assembly on the morning of the 15th, referencing PWF, IAU President Jan H. Oort declared that once again "astronomy has come into conflict with society."[70]

Its Commission 40 on radio astronomy met for two hours on the 18th to consider the SSB conclusions and the Kennedy Administration's statement of U.S. government policy with respect to PWF:

*[A] general discussion ensued which showed that the great majority of the members of the Commission were completely opposed to the West Ford project being attempted under present conditions, not being convinced in particular that the belt of dipoles may have a limited lifetime. This point of view is expressed in a resolution proposed by [Hermann] Bondi, [Wilbur N. "Chris"] Christiansen and [Thomas "Tommy"] Gold, which is adopted unanimously by the Commission.*[71]

A working group consisting of members Lilley, Charles L. Seeger (Stanford University), Francis Graham-Smith, Haddock, Edward F. McClain (Naval Research Laboratory), Andrei P. Molchanov (Soviet Academy of Sciences), Gold and Émile-Jacques Blum (Observatoire de Paris-Meudon) was formed to study the observability of dipole belts.

At a second two-hour meeting on the 21st, Commission 40 adopted a resolution on PWF:

*Commission 40 expresses its appreciation for the fact that the plans for project West Ford have been publicly announced well ahead of launching and that it has been stated officially that the U.S. Government policy on further launchings will be guided by the principle that such projects should not be undertaken unless sufficient safeguards have been obtained against harmful interference with astronomy.*

*Nevertheless, Commission 40 views with the utmost concern the possibility that the belt of dipoles proposed in project West Ford might be permanent, and is completely opposed to such an experiment until this question is clearly settled in published papers and time has been given for their study. Whatever the limitations of present radio*

*astronomical equipment, the Commission is inflexibly opposed to any steps that might permanently compromise future development in radio and optical astronomy.*[72]

Haddock reported back on the outcome of discussions of the working group, suggesting a number of possible experiments that would follow a successful PWF test.

A second resolution was proposed by Seeger and Joseph L. Pawsey (Division of Radiophysics, Council for Scientific and Industrial Research) and adopted by a vote of 7-4:

*If the objections of Resolution 2 above can be removed and the experiment West Ford is performed, Commission 40 regards it as essential that the fullest observations of, or experiments on, the properties and variations of the belt be made by all possible means.*

*Such observations should be made and analyzed according to the highest scientific standards and by means of the best equipment available, bearing in mind that barely detectable signals today may be a great source of interference to future scientific research with more sensitive equipment.*

*These observations or experiments are likely to be difficult to perform and will in many ways parallel those carried out by the bodies responsible for performing the experiment West Ford. Moreover, much specific information such as precise and up to date ephemeris data will be required in any case. Commission 40 therefore urges the establishment of channels in the IAU to obtain fast and full co-operation among the astronomers making such observations and to provide for world-wide dissemination of the results along accepted standards in scientific research.*

*Viewing the position taken by the U.S. Government that any decision on later experiments of this type would be contingent on the results of the analysis of the presently proposed experiment, Commission 40 appreciates the offer of the U.S. Government to extend co-operation and, in particular, asks that the U. S. Government grant full privileges to a group of astronomers, acceptable to the IAU to cooperate with project West Ford authorities to perform quantitative experiments and observations using West Ford facilities with the purpose of determining the properties of the belt and its variations with time and position and to assess its impact on present and future research in astronomy and radio astronomy.*[73]

Lastly, a third resolution proposed by Christiansen (but numbered "1" in the Commission's report) was adopted unanimously by the Commission:

*Commission 40 views with concern the increasing contamination of the space around the Earth by radiating and scattering objects. It feels that no group has the right to change the Earth's environment in any significant way without full international study and agreement.*[74]

The Commission then moved to transmit all three resolutions for consideration by the IAU Executive Committee. The Executive Committee announced a resolution calling for a moratorium on launches of dipole belts until the full implications of their use in terms of impacts to astronomy was understood.

*Maintaining that no group has the right to change the Earth's environment in any significant way without full international study and agreement; the International*

*Astronomical Union gives clear warning of the grave moral and material consequences which could stem from a disregard of the future of astronomical progress, and appeals to all Governments concerned with launching space experiments which could possibly affect astronomical research to consult with the International Astronomical Union before undertaking such experiments and to refrain from launching until it is established beyond doubt that no damage will be done to astronomical research.*[75]

An IAU West Ford Committee was formed of a number of astronomers from both the U.S. and other countries including several members of the SSB committee. Findlay and Donald H. Sadler (Her Majesty's Nautical Almanac Office), then Secretary General of the IAU, were appointed as secretaries.

## *Final U.S. Government position on PWF*

In September 1961 a study of potential interference effects on astronomy conducted by a special PSAC panel chaired by John W. Tukey, a statistician and Princeton professor, concluded that "the United States can proceed with the West Ford communications experiment without danger to science." According to Findlay, "This group ... went somewhat further than previous thinking by saying that in its opinion this conclusion was justified even if the dipoles remained aloft for very many years. The group gave rather deeper study to the possibilities of interference to sensitive radar systems for planetary radar astronomy than had been done previously and found no serious difficulties evident."[76] Wiesner received and approved the panel's findings.

In an 8 September *Science* editorial titled "Two Cheers for West Ford", Associate Editor Joseph Turner praised the U.S. government's handling of PWF with respect to potential impacts to science:

*In evaluating American conduct, however, the basic point to remember is that if our policy had been to run the entire experiment secretly, detection of the belt, according to informed sources, would have been unlikely in the extreme. Lacking advance notice of the project, no one would have been the wiser. The present American approach contrasts favorably with our handling of project Argus, in which, in 1958, without public notice, several atom bombs were exploded at low altitudes to create temporary radiation belts. It also contrasts favorably with the general Soviet practice of announcing launchings as accomplished facts.*[77]

# PWF tests and their aftermath (1961–1964)

The basic design of the project was decided by October 1960, and the copper needles and their dispenser were readied for an anticipated launch before the end of May 1961. An orbit was carefully selected near resonance so as to ensure that the dipoles would fall back to Earth quickly.[78] But the Air Force made changes to the operational requirements of the launch vehicle, precluding it reaching the desired altitude. The Air Force proposed an alternative altitude that would not be resonant, and scientists calculated that the lifetime of the needles in that orbit might exceed 20 years. If the predictions about the dipole belt's impact on astronomy were wrong, they concluded that such a lifetime was unacceptably long

> *since the estimates which had been made of the astronomical effects of the belt had not been able to foretell with accuracy the sensitivity of the observing techniques which might be available 20 or more years hence. The danger was difficult to assess, since, for example, the question of how dark the night sky might be when viewed from a space vehicle above the atmosphere could not be accurately answered, yet within 20 years many orbiting astronomical observatories might be in use at altitudes above most of the atmosphere, yet below the West Ford belt.[79]*

The Committee met and urged the Air Force to delay the launch until the vehicle could reach the requisite altitude, a decision that "must have been most unwelcome to all those who had worked so hard for the first shot."[80]

## *First PWF test launch (October, 1961)*

The first test was conducted when an Atlas-Agena B rocket carrying the MIDAS (MIssile Defense Alarm System) 4 satellite lifted off from Vandenberg Air Force Base at 1353 UTC on 21 October 1961.[81] The PWF dispenser, containing 350 million copper needles with a total mass of 28 kg, rode as a payload aboard MIDAS 4. The needles were embedded in 2.1 kg of naphthalene formed into a cylinder 14 cm in diameter and 32 cm in length. A ground command released the payload into a 3500 × 3750 km orbit at an inclination of 95.9º. Upon release, the cylinder was supposed to spin up to a rotation rate of six revolutions per second. As the naphthalene warmed and melted upon exposure to sunlight, the needles would be ejected more or less uniformly. However, the dispersal mechanism failed. The dispenser did not spin up as planned, and the naphthalene did not melt. The Lincoln Laboratory detected "fragments of the package" with cross-sections of 0.06–0.6 $m^2$ using ground-based UHF radar, finding that they were not sufficiently distributed to relay communications signals.[82]

Likely there were thousands of additional clusters produced in the event too small to be tracked from the ground.[83] A radar image made on 3 November suggested that the canister had indeed separated from MIDAS 4 successfully, but the validity of the image was later called into question after no further detections were made.[84] In 2001, it was estimated that some three-quarters of a million needles from the 1961 test totaling 60 g of mass still remained in orbit as of 1 August 1999.[85]

Lincoln Laboratories decided that the test had been a failure. *Science* reported that MIT was unsure as to whether it would try again:

> *A spokesman at the Lincoln Laboratory said a second attempt would not be made until efforts were completed to account for the failure. It appeared that this would not be accomplished quickly. ... The controversy over whether the project would interfere with radio and optical astronomy has led to considerable caution in the decision on when to attempt another shot. The Lincoln Laboratory wants to be as certain as possible that the second attempt will not be followed by a sudden blossoming of filaments from the first canister.[86]*

On 1 March 1962, the SSB panel circulated "a very full description of what occurred and of the probable reasons for the failure" of the first test launch.[87] The first PWF test also "served as a useful test of the plan set up by the Space Science Board group with Lincoln Laboratory and the Air Force to distribute launch and orbital information as soon as possible after a launch." Before any test that might follow, "there was also plenty of time for the IAU West Ford Committee, and later the IAU Executive Committee, to weigh the question of whether to object to another launch. The outcome was that they decided not to do so."[88] On 8 March, Findlay sent a memo to the IAU West Ford Committee on behalf of the SSB summarizing the results of the launch and making data from the test available to the community.[89]

The United Nation Committee on the Peaceful Uses of Outer Space (COPUOS), created by the General Assembly in 1958, met in March 1962 for the first time since differences of opinion between the U.S. and U.S.S.R. concerning the Committee's administration were resolved. Project West Ford was on the agenda. In a position paper submitted to the proceedings, the U.S delegation summarized events to date, speculated on an "anticipated foreign position", and articulated the "United States Position". It included a tacit admission that the international astronomical community had received news about PWF poorly:

> *Some foreign scientists have expressed concern about possible interference of the experiment with optical and radio astronomy, if it is longer-lived than estimated. These expressions of concern were made both before and after the launch of the first WEST FORD package. ... The experiment has been discussed more thoroughly in advance with a broad range of scientists of other countries than any other space experiment. If certain astronomers of other countries remain concerned, we regret it, but the extensive discussions which have been held do not, in our view, develop cause for such concern. It should be noted that at no time has any other government expressed official concern to the United States.*[90]

The paper reiterated the main points of the President's Policy Statement of 8 August 1961, including a commitment by the U.S. that "no additional launches of orbiting dipoles will be planned" until the outcome of the first test launch was "analyzed and evaluated and that findings of foreign as well as United States scientists will be taken into consideration."

Armed with engineering data on the test, members of a working party within the Royal Society Steering Group on Space Research, chaired by Ratcliffe, published its findings on the possible radio and optical astronomy impacts of a successful deployment of needles. The radio astronomy report, led by A.C. Bernard Lovell (University of Manchester) and Martin Ryle (University of Cambridge), concluded that "even with existing techniques, and with the density of dipoles planned for the West Ford test belt, radio or radar transmitters could cause limitations to observations made with large radio telescopes." Yet they did not believe the probability of catastrophic interference was high:

> *For densities such as were planned for the West Ford test belt, and for small radio telescopes used with existing receivers, the interference is only likely to be severe when the main beam or near-in side lobes of the radio telescope intersect the belt at the same point as does the main beam of a typical transmitter. The probability of this is likely to be very low in circumstances likely to arise in the near future.*[91]

Still they used the occasion to call for globally protected radio frequency allocations set aside for radio astronomy: "[U]nless permanent world-wide frequency bands are allocated exclusively for radio astronomy, there is a severe danger of interference from remote transmitters scattered from belts of orbiting dipoles."

The authors of the optical astronomy report, Donald E. Blackwell (University of Oxford) and Robert Wilson (Atomic Energy Research Establishment), suggested that previous estimates of the dipole belt brightness did not consider "the full properties of the scattered sunlight, particularly with regard to polarization effects, and have usually compared the intensity of scattered light with the general sky background."[92] They modeled the monochromatic surface brightness of the PWF dipole belt, predicting it would be $2.7 \times 10^{-9}$ erg cm$^{-2}$ Å$^{-1}$ s$^{-1}$ sr$^{-1}$ at 500 nm. They compared that to the minimum night sky brightness from all natural sources (night airglow near solar minimum, scattered starlight and the zodiacal light), which they quoted as $1.3 \times 10^{-7}$ erg cm$^{-2}$ Å$^{-1}$ s$^{-1}$ sr$^{-1}$ at the same wavelength. The dipole belt would, in their determination, therefore contribute about 2% excess light above the natural night sky background.

Introducing the papers, RAS Secretary Hermann Bondi noted the "considerable divergences of opinion both on the effect of a test belt and of any denser operational belt, and also about the permanent or temporary nature of such orbiting dipoles." Bondi felt that the campaign to raise concerns about PWF had been at least partially successful:

*[It] is certainly true to say that to the best of its ability the Society has added its voice to those of many other bodies, national and international, who are trying to prevent any interference with the future development of astronomy. We have persistently laid the greatest stress on the need to make certain that any belt put up would at least be temporary, since a permanent damaging of the prospects of astronomy would be a very serious matter indeed.*[93]

Faced with the ongoing criticism, PSAC panel member and Harvard physicist Edward M. Purcell took to the pages of *New Scientist*, painting the objections of astronomers to PWF as fundamentally irrational. Reaction to the experiment, he suggested, was "sharp and vocal ... most conspicuously in Britain."

*Some of it came from scientists who had been kept fully informed of the prior studies, but who had wanted all along to stop the experiment in order to forestall any future activity of the sort. Some were even unwilling to grant the experiment a valid purpose, and through their assertions runs the implication that the enterprise was essentially frivolous, the brain-child of irresponsible "military engineers". To the extent that this reflects the weariness, even revulsion, that many of us feel, as scientists, at the interminable preoccupation of technology with military problems — to that extent I understand the feeling.*

But Purcell also detected a "larger question" in the PWF experiment:

*In our time, technology has brought about irreversible environmental change on a large scale. It will bring about others, intentionally or otherwise. ... Even desirable changes will seldom be manifestly benign in every aspect. Shall we take charge of our affairs like rational men? If objective balancing of risks can never justify an irreversible action, responsible action is simply blocked, while irresponsible action remains unchecked. My Government had a valid reason for wanting to distribute 75*

*lb of copper through 10 million cubic miles of space. It brought the question into the open a year ahead of time for quantitative examination by the world scientific community. The action taken was, in my view, responsible and rational. I suspect that the establishment of such a precedent will prove to be the most significant result of West Ford, whatever the future may hold for that mode of radio communication.*[94]

### *West Ford Drag (April 1962)*

Despite its public representation at the U.N. COPUOS meeting the month before that it "[did] not plan to consider the question of additional belts until results of the first have been analyzed and evaluated," a successful test of more limited scope was conducted that involved a much smaller number of larger dipoles. Dubbed "West Ford Drag," the payload was launched with the MIDAS 5 satellite on an Atlas Agena from Vandenburg at 1504 UTC on 9 April.[95] The spacecraft reached a nearly circular 3000 km orbit, but the satellite failed the next day. Before failure, it successfully ejected an experimental package consisting of a fiberglass cylinder that contained six large copper dipoles designed to test the theory that interaction with the magnetosphere would induce a charge on the dipoles, yielding a drag force causing their orbits to decay rapidly. The dipoles were made sufficiently large to ensure that they could be tracked individually from the ground. The experiment proved that "charge drag" was not a significant effect, providing an important constraint on the expected lifetime of the PWF needles.[96]

### *Second PWF test launch (May 1963)*

The geopolitical circumstances of the development of PWF continued their own parallel evolution. The U.S.S.R. abruptly ended its self-imposed moratorium on atmospheric nuclear tests in August 1961; in response, the U.S. quickly ramped up plans for its own series of five high-altitude tests known as Operation Fishbowl.[97] After two failed test launches in June 1962, it succeeded spectacularly on 9 July with Starfish Prime,[98] which became the largest nuclear test ever conducted in outer space. The detonation generated an electromagnetic pulse far stronger than predicted, interfering with electronics on the ground hundreds of kilometers from the launch site. Starfish Prime also damaged or rendered permanently useless one-third of all active communications satellites then in orbit.[99] The results made all the more real the original fears that precipitated PWF, adding exigency to the Air Force's plans.[100]

Meanwhile, in May 1962 the IAU West Ford Committee decided to back down from its earlier (and total) opposition. Faced with results from the first test launch, the committee felt satisfied that a successful deployment would not cause immediate harm to either optical or radio astronomy. Instead, it took a cautious, wait-and-see stance and committed to cooperate with Lincoln Laboratory and the SSB to collect and share data on the resulting dipole belt. This did not sit well with some astronomers, who continued to complain to the popular press that the Air Force did not engage in adequate consultation with them and further that launching additional dipole belts would contribute to the pollution of the space environment. The run-up to the second PWF test took place against the backdrop of the 1962 publication of Rachel Carson's iconic *Silent Spring* and the public's growing anxiety

about both "unseen pollutants" and an unprecedented number of atmospheric nuclear tests.[101]

On 3 May 1963 the SSB sent a memorandum to the IAU West Ford Committee and to volunteer observers and observatories throughout the world to the effect that another PWF test would be made "in the near future"; the launch was announced to the public three days later. At 2006 UTC on 9 May, an Atlas Agena B lifted off from Vandenburg carrying the MIDAS 6 spacecraft and the second PWF test package.[102] For this test, 480 million needles, each 1.8 cm long and 0.0018 cm in diameter, were embedded in 1.5 kg of naphthalene in the form of five disks and loaded into the dispenser. Engineers added a telemetry transmitter to monitor the spin rate of the dispenser.[103]

At 2357 UTC on 10 May, ground controllers commanded the ejection of the canister with a spin rate of eight revolutions per second. The naphthalene disks entered a 3600 × 3680 km orbit at an inclination of 87.4°. This time, uneven solar heating resulted in partial dispersal of the needles. Shapiro recalled that "the separation of the dipole canister from the satellite was delayed by [about] 20 minutes, for reasons over which we had no control, causing the sunlight not to strike the canister at the correct angle for proper dipole dispensing." (Shapiro 2023) Somewhere between 15–40% of the dipoles were freed from the naphthalene and dispersed, while the remaining mass remained in clumps of varying sizes.[104]

The dipoles were first observed in orbit by the Lincoln Laboratory radar on the 12th, and by the next day the dipole cloud reached a linear extent of about 7500 km along its orbit.[105] On 15 May, two-way communications using the dipole belt as a passive reflector were successfully demonstrated. (Figure 2) Near the end of the month, Findlay began his first search for 7.75 GHz signals from the Millstone Hill transmitter scattered by the dipole belt to the 85-foot Howard E. Tatel radio telescope at NRAO. The campaign ran until 7 June.[106] Both it and a second campaign (14–21 October) yielded negative results.[107]

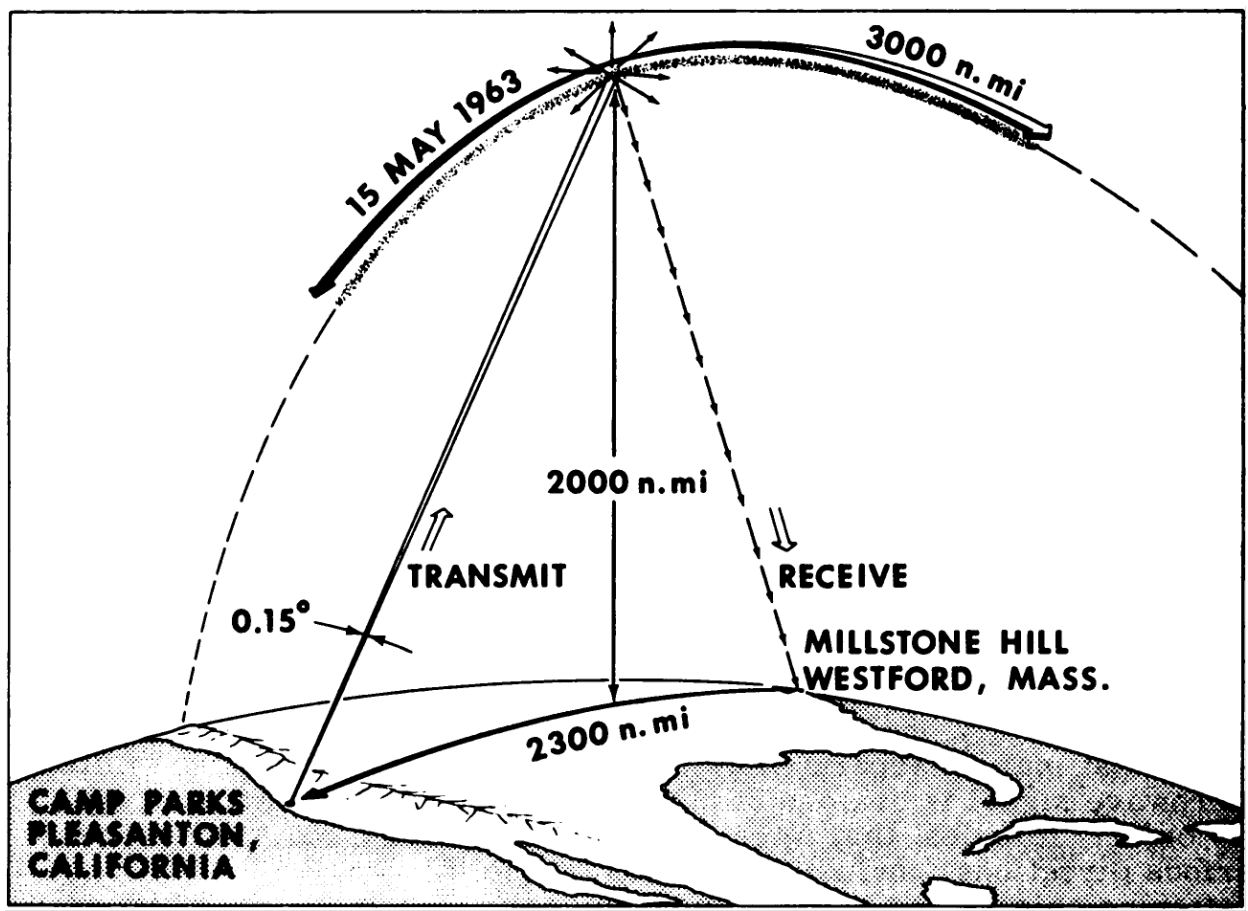

**Figure 2**. A schematic illustration of radio signals relayed from the Project West Ford dipole belt after the May 1963 test launch.[108]

By mid-June, some 40 days after launch, the needles that escaped the naphthalene disks had dispersed sufficiently to form a closed toroidal belt around the Earth. Radar observations suggested that the belt had a circumference of 36000 km at a mean altitude of 3650 km, and cross-sectional dimensions of 15 × 30 km. The volume density of the needles was therefore between three and nine needles per cubic kilometer of space, close to the expected value. The dipoles had a predicted an orbital lifetime of less than seven years.[109]

Astronomers began searching for optical and radio signals attributable to the dipole belt within days of its deployment. A photometric campaign was conducted by Tifft *et al.* between 20 May and 21 June that claimed to have "clearly detected" the belt at a brightness equivalent to 0.5% of the total night sky brightness in the same direction.[110] Shapiro et al. wrote in reaction that "most of the conclusions drawn by Tifft, et al. from their observations seem unwarranted," arguing that, in fact, there were no dipoles in the direction of the part of the sky they observed and that their alleged detections were nothing other than "sky noise."[111] However, they praised the "excellent photoelectric observations" of Palomar Observatory astronomers Alan Sandage and Charles Kowal, who found an upper limit to the brightness of the dipole belt of about 4% of the night sky brightness.[112]

Measurements continued during May and into June, by which time it was assumed that the dipole belt had reached its greatest spatial extent. As they awaited results of the observing campaign, some astronomers continued to express deep dismay about PWF. For example, on 24 May, the Harvard Crimson quoted Lovell: "The damage lies not with this experiment alone, but with the attitude of mind which makes it possible without international agreement and safeguards." Other scientists stated publicly that the way the project was actually carried out "might be a satisfactory precedent for similar future discussions and decisions."[113] Sandage and Kowal wrote that while their photoelectric observations of the dipole belt suggested that the PWF test belt was "harmless to ground-based astronomy,"

> *future experiments of a similar nature with a larger payload may not be so lucky. If, for example, a West Ford operational belt is established with, say, 100 times the present payload, the belt would be four times brighter (shortly after launch) than the natural night sky radiation, a level which would begin to be serious for certain types of astronomical observations. Once the principle of unilateral contamination is accepted, the doors are opened for disastrous future possibilities and man may well succeed in changing his environment (albeit astronomical) beyond repair.*[114]

Concern over the effects of PWF were raised at the United Nations in June; the U.S. ambassador to the U.N., Adlai E. Stevenson, defended the project.[115]

In December the Findlay Committee submitted its final report to the SSB. It identified two major findings:

> *(a) We conclude, as was originally forecast, that as of the time of this statement and with the observing techniques in use today the present West Ford experiment has not been harmful to either optical or radio astronomy.*

> *(b) The predictions of the effects of the experiment have been reasonably well borne out by the observations. We may therefore rely on essentially the same methods to predict the effects of any experiments similar to the West Ford experiment which may possibly be proposed in the future, suitable allowance being made for the increased vulnerability that may be associated with future advances in observing techniques.*[116]

The report was made public on 26 March 1964.[117] Liller published the results of the observing campaign in January 1964. The data showed that the optical brightness of the belt formed in the second PWF test was consistent with predictions (Figure 3) and lower than first anticipated by Lincoln Laboratory, particularly in the first month after ejection. As the needles dispersed and the belt grew in physical extent, the volume density of dipoles and the apparent brightness of the belt both quickly decreased. By the time of the last firm optical detection of the belt, at Palomar Observatory on 18 May 1963, its surface brightness was roughly 1.6% that of the night sky. Non-detections as late as 30 May confirmed an upper limit of about half that figure.

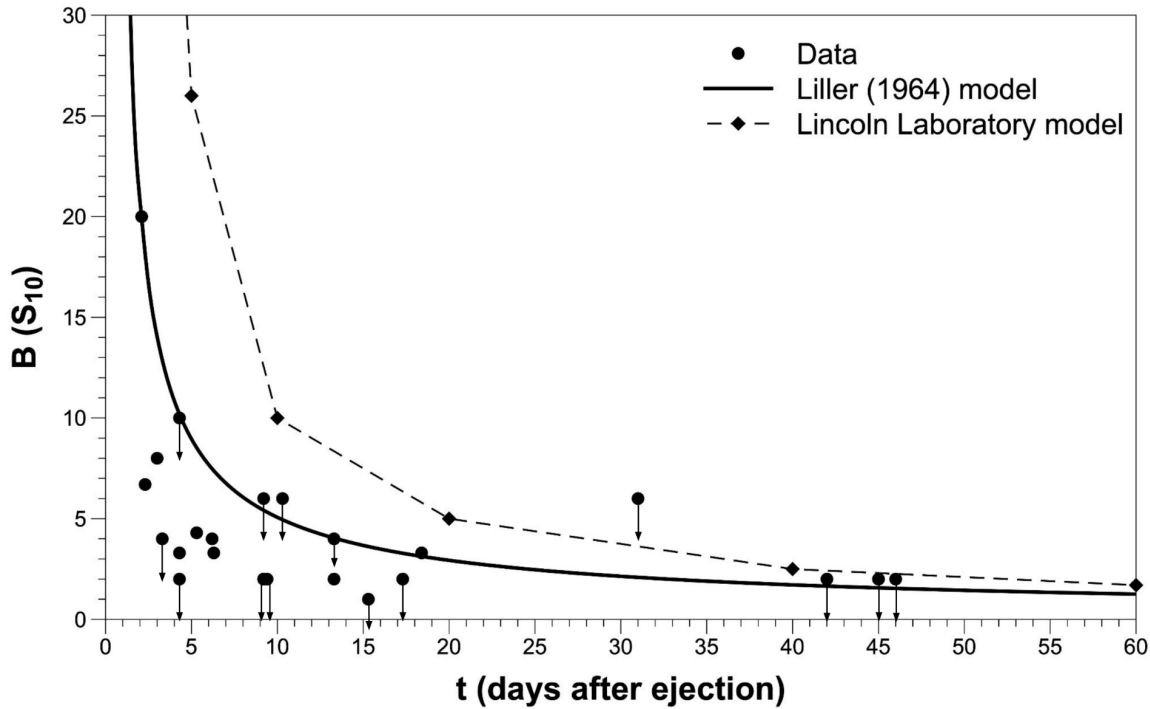

**Figure 3**. Measured optical surface brightness of the dipole belt dispensed in the second PWF of May 1963 (solid circles)[118]. The units on the ordinate are the number of tenth magnitude stars per square degree ($S_{10}$), a linear brightness unit; for reference, a pristine night sky under quiescent airglow conditions has a zenith brightness of about 205 $S_{10}$, which is about $2\times10^{-4}$ candela per square meter.[119] Ejection of the dipoles is assumed to have occurred on UTC 1963 May 11.0. The solid diamonds joined by the dashed line are pre-launch predictions by Lincoln Laboratory, quoted by Liller, and the solid line is Liller's prediction. Pre-launch predictions for the steady state at $t$ = 60 days by Morrow and MacLellan[120] and Blackwell and Wilson[121] are within the size of the model points plotted there and thus do not appear. Uncertainty estimates on the plotted observations are not quoted in the source material.

Given this outcome, Findlay summarized what he felt were statements of fact concerning the experiment:

*1) The U.S. Government had a valid reason for wishing to conduct the experiment.*

*2) The plan of the experiment was placed before the world scientific community more than two years before the final launch.*

*3) The assessments before the launch of the effects of West Ford on science were soundly based and soundly made.*

*4) The decision to carry out the experiment under formally stated policy conditions was responsible and rational.*[122]

In November COSPAR acknowledged the findings of no significant impacts on optical or radio astronomy resulting from the PWF test, "and recommends to its Members that any proposals for future experiments of this sort also be given the benefit of thorough evaluation by the scientific community and notably by the International Astronomical Union, in order to check in advance their harmlessness to other scientific research."[123] The following May, an unclassified report to the Defense Science Board concluded "there has been general agreement that no interference to optical or radio astronomy has resulted from the original West Ford experiment in May 1963."[124] The report insisted that the concept tested by PWF was still viable, but found that the continued maturity of missile launch capabilities and more robust electronic equipment aboard satellites made dipole belts as passive microwave reflectors less effective than satellite relay systems. Technology had obviated the need to deploy an 'artificial ionosphere,' and the U.S. government considered the matter concluded.

Some questions remained about the fate of the needles in the second PWF test. Shapiro and Jones predicted in October 1961 that the dipoles should re-enter in "about 7 years". In December 1966 Shapiro wrote that "the dipoles reentered the dense lower atmosphere during a period of several months, centering about the predicted reentry date of 1 January 1966 for the average dipole."[125]

*Unfortunately, New Year's celebrations were not punctuated by a display of tiny fireballs. Calculations show that because of their high A/M* [area to mass] *value and shallow reentry angle, the dipoles were able to radiate heat rapidly enough to avoid disintegration, and most probably floated back to Earth essentially unharmed. Even had they disintegrated during reentry, the dipoles would have produced trails far too faint to have been visible.*[126]

He quoted an expected surface area density of about five dipoles per square kilometer in the polar regions of Earth, suggesting a vanishingly small probability that anyone would ever find them.[127]

"This report on the fate of the dipoles is intended to be the last," Shapiro wrote as he closed his article in Science.

*Our gratification at having seen experimental results in agreement with predictions is tempered by the realization that little can be done to clear the clouded reputation of Project West Ford. For, as was observed long ago, the (alleged) evil that projects do lives after them; the good is oft interred with their bones. So let it be with West Ford.*[128]

## West Ford and the development of the Outer Space Treaty (1962-1967)

Project West Ford was hardly the only concern prompted by the rapid pace of human activities in outer space from 1957. Potential military uses of space, including the possibility of using intercontinental ballistic missiles transiting through space to deliver nuclear weapons, caught the attention of scientists and diplomats alike. In August 1957 the United States suggested a regime for the verification of the testing of space objects before the launch of the first artificial satellite, followed by a proposed inspection system.[129] The Soviet Union, by then nearly on the cusp of the launch of *Sputnik 1*, rejected these overtures.

At the meeting of COPUOS in March 1962, the U.S. acknowledged that PWF could be considered a military action in space: "It is possible that charges may be made that the United States is 'contaminating' outer Space, that we are unilaterally (and in the face of objections by foreign scientists) performing experiments which interfere with the peaceful scientific pursuits of others, and, since the experiment is sponsored by the United States Air Force, that we are using space for military purposes."[130] While the U.S. had "consulted foreign scientists extensively" in the run-up to the first test, it categorically ruled out any possibility that "consultation with and approval by other countries are prerequisites to conduct our space activities." It did so because otherwise "such commitments would not be compatible with our view that outer space is freely available for exploration and use consistent with the United Nations Charter and other international law and agreements and would place presently unacceptable limits on our freedom of action."

Along with the events of PWF, the Antarctic Treaty (1959) modeled a potential off-ramp from an otherwise pending militarization of space.[131] It provided a framework for the governance of the only continent lacking a permanent human population, and represents the first arms-control agreement of the Cold War.[132] The Treaty designated the whole of Antarctica as a scientific preserve and prohibited any future military activity on the continent. Twelve countries that maintained an active presence in Antarctica for the IGY, including both the U.S. and the U.S.S.R., were the original signatories to the Treaty. There were clear parallels between Antarctica and outer space, both being a legal 'terra nullius' belonging to no one.[133] As tensions between the Soviet Union and the West rose, so too did fears that a mistake made in either province would quickly lead to war. U.S. President Dwight D. Eisenhower put this into words during his address at the U.N. General Assembly on 22 September 1960, during which he formally proposed that the principles of the Antarctic Treaty be applied to govern activities in outer space and on Solar System bodies beyond the Earth.[134]

The United States and Soviet Union were unable to resolve their differences on many of the details having to do with conducting space activities. Neither was willing to compromise on any point that it determined would weaken its national security. In May 1963, the Scientific and Technical Subcommittee of COPUOS recommended that the Committee put its full attention on the possibility of harmful interference with the peaceful uses of outer space.[135]

The Soviet posture on space disarmament suddenly shifted with the signing of the Treaty Banning Nuclear Weapon Tests in the Atmosphere, in Outer Space and Under Water on 5 August 1963. On 17 October, the U.N. General Assembly unanimously adopted a resolution that would prohibit the deployment of weapons of mass destruction in space.[136] The following month, COPUOS endorsed without change a draft resolution negotiated by the U.S. and U.S.S.R. that shortly thereafter was adopted by the General Assembly as the Declaration of Legal Principles Governing the Activities of States in the Exploration and Uses of Outer Space. Its Principle 6 recognized efforts like PWF that "would cause potentially harmful interference with activities in the peaceful exploration and use of outer space" should be made subject to obligatory consultations between countries engaging in such activities and those that might be adversely affected:

> *In the exploration and use of outer space, States shall be guided by the principle of cooperation and mutual assistance and shall conduct all their activities in outer space with due regard for the corresponding interests of other States. If a State has reason to believe that an outer space activity or experiment planned by it or its nationals would cause potentially harmful interference with activities of other States in the peaceful exploration and use of outer space, it shall undertake appropriate international consultations before proceeding with any such activity or experiment. A State which has reason to believe that an outer space activity or experiment planned by another State would cause potentially harmful interference with activities in the peaceful exploration*

*and use of outer space may request consultation concerning the activity or experiment.*[137]

Further negotiations toward arms control in outer space continued over the following three years. Various proposals were discussed during the General Assembly session in December 1966, including competing draft treaties that both the U.S. and U.S.S.R. submitted in June of that year.[138] At meetings held in Geneva during September, the two countries reached agreement on the main provisions of what became the Outer Space Treaty.[139] The General Assembly approved by acclamation a resolution endorsing the Treaty, which was opened for signature on 27 January 1967. Having been signed by the requisite number of state parties, the Treaty entered into force on 10 October 1967. Its Article IX draws its inspiration from Principle 6 of the 1963 principles declaration, which in turn traces back to astronomers' objections to PWF in part on the basis that it directly contradicted the notion that space activities should be carried out by countries with consideration for what effects they might yield on other countries.

## Applying the lessons of PWF to understand modern uses of outer space as they impact astronomy

Project West Ford was successful in the sense that it demonstrated the feasibility of carrying out radio communications over long distances by a method that involved little risk and yet was resilient in the face of destructive jamming methods. It did not appear to yield the harms to optical or radio astronomy that its detractors feared. Some astronomers were worried about specific impacts that turned out to be unfounded, while others, some in alignment with the interests of the U.S. government, decided their continued protests were irrational in the face of modeling results suggesting that the dipole belt would have no impact.

Although it was a fully worked proof of concept, the timing of PWF virtually assured that it would see no further technological development. Shortly after the second test launch true two-way radio relays in space through active satellite technology became available, obviating the need for passive reflector systems. Yet its lasting contribution to the history of astronomy, which echoes down to the present, is that it provoked the first debates about the 'environment' of outer space and how it could be negatively impacted by human activities taking place there. In the end, despite its self-awareness, PWF prompted the enactment of certain environmentally sensitive provisions of international outer space law. This informed the process by which the Outer Space Treaty was framed and continues to be implemented and interpreted.

Scientists continued to assess the PWF affair for years after the tests were concluded. In 1977, Bernard Lovell identified PWF as a "danger" and a cautionary tale in the sense of it as a potential "hindrance to astronomical observations which may result from modern defence R & D activities".[140] An attitude typical of those reflecting on the episode is summarized in the cartoon that accompanied Jean-Claude Pecker's and Gisèle Ringeard's 1973 paper "Astronomy, the People and the Governments." (Figure 4) The PWF dipole belt is shown positioned between an astronomer's "large telescope" and the "secrets of the origin of the universe (hidden)". Nearby sits a pile of "unwritten books, where the secrets of the universe were to be written." The astronomer, described as "desperate," is "trying hard to understand the origin of the universe and get the Nobel prize before the Westford engineers." The cartoon represents the anxieties of astronomers both before the PWF tests, as well as in their aftermath while the U.S. government still pondered whether to launch a larger, operational version.

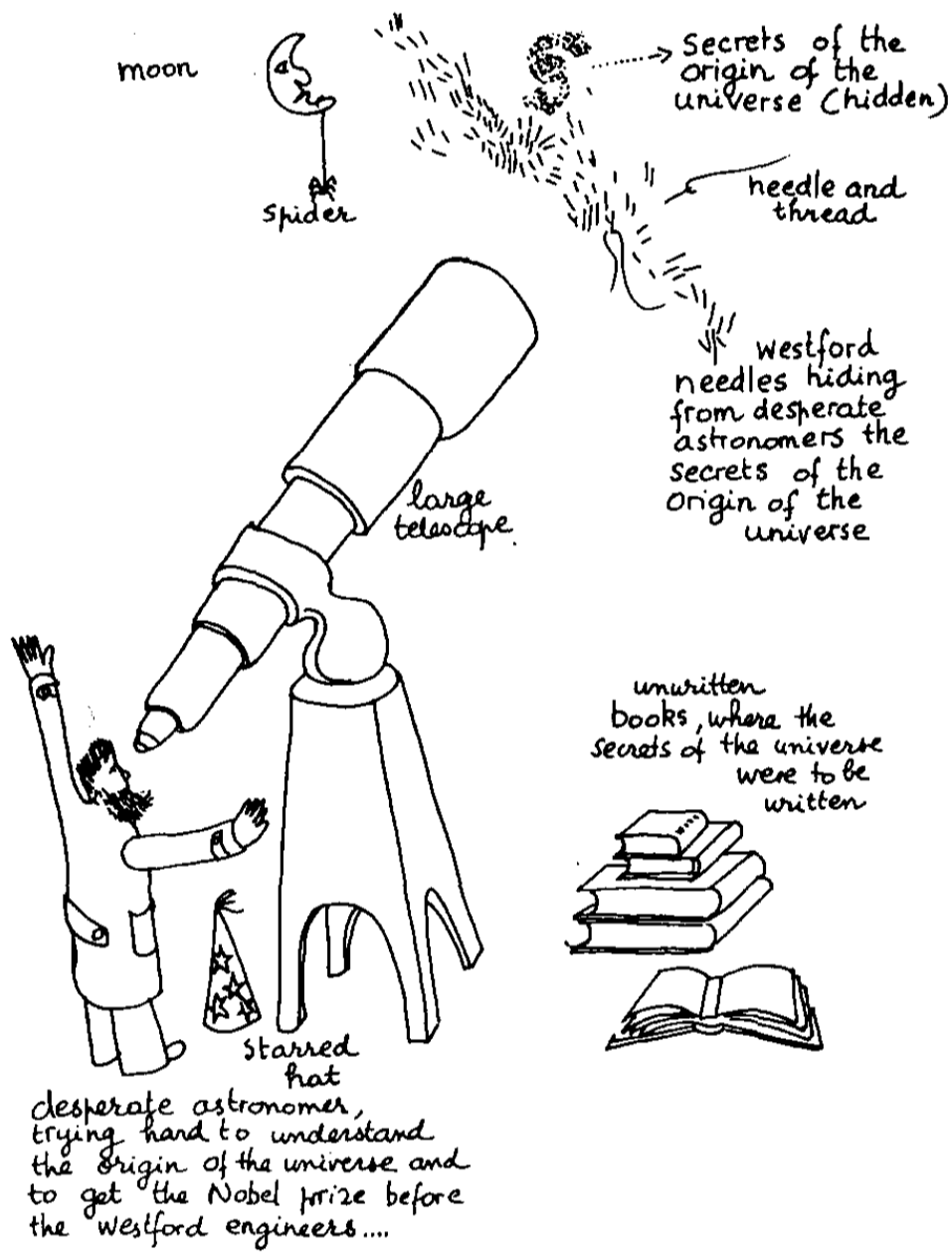

**Figure 4**. A 1973 cartoon poking fun at the Project West Ford affair from the perspective of astronomers.[141]

The role of astronomers in defining an emerging sense of "space environmentalism" is an integral part of the PWF story that brings its lessons forward into the modern era. PWF scholar Lisa Ruth Rand argues that it was "attempts by astronomers in America and Europe to appoint themselves as international environmental watchdogs for outer space"[142] that ultimately situated them "as the appropriate stewards to provide the necessary environmental oversight."[143] Six decades on from the events of the early 1960s, Michael Byers and Aaron

Boley observe that "it is difficult to imagine a clearer confirmation that the term 'exploration and use of outer space' in its international law context has therefore always included astronomy and continues to do so."[144]

Satellites are an inextricable part of this story, whether as the dispensers of the PWF needles or craft like early satellites in the U.S. *Explorer* series, which ironically proved the existence of the Van Allen belts,[145] blurring the line of demarcation between the ionosphere and outer space. It is also arguable that the nuclear tests that disrupted the ionosphere (against which PWF was a reaction) were reckless and not peaceful in nature. While PWF had military origins, it was carried out in a very different way.

In our own time, many of the same concerns raised by astronomers during PWF have appeared again in connection with satellites.[146] The ongoing deployment of satellites, which may number in the tens of thousands by 2030, has again caused anxiety about their effects on the night sky. As they transit through the fields of view of science instruments attached to telescopes, they contribute optical and infrared light that can impair the detection of faint light from astrophysical sources.[147] Space telescopes are not immune to this interference.[148] Orbital crowding that leads to rapid increases in space debris production may raise the diffuse brightness of the night sky.[149] Mitigating the effects of satellites on astronomy is expected to increase the costs of operating large telescopes and lengthen the time required to complete major projects such as the Vera Rubin Observatory's Legacy Survey of Space and Time.[150] Radio astronomy is affected by both direct and indirect radio-frequency interference from satellites.[151]

There are also concerns about the negative effects of the life cycle of satellites generally on the Earth-space environment. These include the possibility that debris from a large flux of re-entering satellites could create a cloud of plasma around the Earth that may yield a perturbation on the magnetosphere.[152] Metal atoms from re-entering spacecraft exceeding the natural background flux from micrometeoroids are observed in stratospheric aerosol particles,[153] which could change the thermal balance of the atmosphere akin to proposed 'geoengineering' efforts.[154]

Beginning in 2020, astronomers raised global attention to these issues through research, advocacy and dialog with key players in the commercial space industry.[155] In 2022 the International Astronomical Union formed the Centre for the Protection of the Dark and Quiet Sky from Satellite Constellation Interference (IAU CPS), which has called for industry and government adoption of certain recommended migitations.[156] Part of their argument rests on the claim that astronomy is entitled to "due regard" on the part of satellite operators when their activities cause harmful interference.[157] They continue to press their case at the United Nations Committee on the Peaceful Uses of Outer Space.[158]

This story is playing out against a background in which some scholars argue there is little meaningful legal regulation of commercial activities in outer space that affect astronomy.[159] As the number of artificial objects orbiting our planet increases quickly, Aparna Venkatesan and co-authors question whether the Outer Space Treaty remains fit for purpose:

> *The use of celestial bodies and outer space for peaceful purposes, avoiding colonial competition and environmental disruption, and discouraging militarization and conflict in space are a few of the main premises of the OST, as well as the (far less widely ratified) Moon treaty. Due to the non-binding nature of international law and lack of oversight or enforcement of the treaty, as well as the broad definitions of key components (such as*

*'colonial competition', 'militarization' and 'peaceful purposes'), it is easy to work around such premises through strategic loopholes. ... Furthermore, the OST was created as the golden era of the space age was just beginning, when competitive ambition was high, resources on Earth and in space seemed limitless, and climate change and environmental concerns were factors on a far horizon. Today—in the language of the current COVID-19 pandemic—we have very different pre-existing conditions, making us deeply vulnerable to the many existential threats facing humanity, all species, and the planet itself.*[160]

While stopping short of calling for a legal instrument to supersede the Outer Space Treaty, Venkatesan *et al*. note that "we are beginning to see (and will probably see far more) examples of the main components of the OST being ignored or actively worked around" and call for "a radical shift in the policy framework of international regulatory bodies towards the view of space as an ancestral global commons that contains the heritage and future of humanity's scientific and cultural practices."[161]

Although there are some similarities between PWF and the satellite 'megaconstellation' phenomenon, there are also distinct differences. The modern problem involves predominantly private, non-state actors whose activities are in principle supervised by states following their obligations under the Outer Space Treaty. While PWF initially developed in military secrecy, the U.S. Government decided to declassify key pieces of information about the project in order to head off criticism that the dipole belts were some kind of hostile action in space. On the other hand, many of the details of spacecraft design that may have bearing on their visibility to telescopes are protected trade secrets not disclosed to the public. Yet in both situations, the initial analysis indicated risks to both spacecraft and astronomy. In the case of megaconstellations, the rest of the story has yet to be written.

## Acknowledgements

The author wishes to thank a number of people for helpful discussions on what became this manuscript. Irwin Shapiro (Harvard University) graciously shared his firsthand recollections about Project West Ford and answering many questions. Cris van Eijk (Newcastle University) provided crucial information tying PWF to the genesis of what became the Outer Space Treaty. Lisa Ruth Rand (California Institute of Technology) offered useful criticism of some of the points made in the presentation. Lastly, the careful, critical reading of the manuscript by two anonymous referees resulted in suggested changes that helped improve the clarity of the presentation and cleared up confusion around certain factual assertions.

## Notes on Contributor

John Barentine is the Principal Consultant and Executive Officer at Dark Sky Consulting, LLC. He earned a Ph.D. in astronomy from the University of Texas at Austin, USA. He is a member of the American Astronomical Society and the International Astronomical Union, and is a Fellow of the Royal Astronomical Society. He co-leads the Community Engagement section of the IAU Centre for the Protection of the Dark and Quiet Sky from Satellite Constellation Interference and is the author of three books on the history of astronomy.

# Notes

1. R. Massey, S. Lucatello and P. Benvenuti, "The challenge of satellite megaconstellations", *Nature Astronomy,* 4 (2020), pp. 1022-1023; O.R. Hainaut and A.P. Williams, "Impact of satellite constellations on astronomical observations with ESO telescopes in the visible and infrared domains", *Astronomy & Astrophysics,* 636 (2020), art. A121.
2. I. Levchenko, S. Xu, Y.L. Wu, and K. Bazaka, "Hopes and concerns for astronomy of satellite constellations", *Nature Astronomy*, 4 (2020), pp. 1012–1014; P. Hasan, "Dark skies and bright satellites", *Resonance*, 28 (2023), pp. 547–565.
3. *New York Times*, 15 May 1921, 1.
4. S. Attwood, "Radio-wave propagation between World Wars I and II", *Proceedings of the IEEE*, 50 (1962), pp. 688–691.
5. V. Bush, *Science–the endless frontier: a report to the president on a program for postwar scientific research* (Washington, D.C.: United States Government Printing Office, 1945).
6. S.W. Leslie, *The military-industrial-academic complex at MIT and Stanford* (New York, NY: Columbia University Press, 1993).
7. H.P. Williams, "The effect of high-altitude nuclear explosions on radio communication", *IEEE Transactions on Military Electronics*, 6 (1962), pp. 326–338.
8. "A surprise to the public, the launch of Sputnik was not unexpected by the U.S. military." J. West, "The Sputnik moment re-examined", *The Ploughshares Monitor*, 42 (2021), pp. 21. See also K. McQuaid, "Sputnik Reconsidered: Image and Reality in the Early Space Age", *Canadian Review of American Studies*, 37 (2007), pp. 371–401.
9. D. Healey, "The Sputnik and Western defence", *International Affairs*, 34 (1958), pp. 145–156.
10. F. McDonald and J. Naugle, "Discovering Earth's Radiation Belts: Remembering Explorer 1 and 3", *Eos: Transactions American Geophysical Union*, 89 (2011), pp. 361–364.
11. L.J. Lanzerotti and D.N. Baker, "International Geophysical Year: Space Weather Impacts in February 1958", *Space Weather*, 16 (2018), pp. 775–776.
12. The electronics aboard early satellites were relatively fragile and not hardened against high fluxes of charged particles during solar geomagnetic events. For contemporaneous concerns, see, e.g., H.E. Newell and J.E. Naugle, "Radiation Environment in Space: Satellites and space probes are revealing the kinds and amounts of radiation men will encounter in space", *Science*, 132 (1960), pp. 1465–1472.
13. H. Gavaghan, *Of Moons and Balloons* (New York: Springer New York, 1998), 180–187.
14. J. Pierce and R. Kompfner, "Transoceanic communication by means of satellites", *Proceedings of the IEEE*, 47 (1959), pp. 372–380.
15. W.K. Victor and R. Stevens, "The Role of the Jet Propulsion Laboratory in Project Echo, *IRE Transactions on Space Electronics and Telemetry*, SET-7 (1961), pp. 20–28.
16. W.C. Jakes, "Participation of bell telephone laboratories in Project Echo and experimental results", *Bell System Technical Journal*, 40 (1961), pp. 975–1028.
17. A. Wilson, "A History of Balloon Satellites", *Journal of the British Interplanetary Society*, 34 (1981), pp. 10-22.
18. W.W. Ward and F.W. Floyd, "Thirty years of research and development in space communications at Lincoln Laboratory", *The Lincoln Laboratory Journal*, 2 (1989), pp. 5–34.
19. J.A.V. Allen, C.E. McIlwain and G.H. Ludwig, "Satellite observations of electrons artificially injected into the geomagnetic field", *Proceedings of the National Academy of Sciences*, 45 (1959), pp. 1152–1171; N.C. Christofilos, "The Argus experiment", *Journal of Geophysical Research*, 64 (1959), pp. 869–875.
20. C. Jones, M. Doyle, L. Berkhouse, F. Calhoun and E. Martin, *Operation ARGUS 1958, Technical Report DNA 6039F* (Washington, D.C.: U.S. Defense Nuclear Agency, 1958).
21. L.M. Mundey, "The civilianization of a nuclear weapon effects test: Operation ARGUS", *Historical Studies in the Natural Sciences*, 42 (2012), pp. 283–321.
22. C. Overhage and W. Radford, "The Lincoln laboratory West Ford program—an historical perspective", *Proceedings of the IEEE*, 52 (1964), pp. 452–454.
23. T.J. Levin, *Contaminating space: Project West Ford and scientific communities, 1958-1965* (Fairbanks, Alaska: University of Alaska Fairbanks, 2000).
24. Levin, *op. cit.* (Note 23), p. 8, footnote 10.
25. I.I. Shapiro, private communication, 29 April 2023.
26. R. Buderi, *The invention that changed the world: how a small group of radar pioneers won the Second World War and launched a technological revolution* (New York, NY: Simon & Schuster, 1997).

27  Figure 3 from *Project West Ford: A feasibility study of orbital scatter communication, by the Massachusetts Institute of Technology, Lincoln Laboratory, Lexington, Massachusetts, March 27, 1964. Satellite Communications, 1964: Hearings Before a Subcommittee of the Committee on Government Operations, House of Representatives, Eighty-eighth Congress, Second Session, Volume 74-77* (Washington, D.C.: U.S. Government Printing Office, 1964), p. 554.
28  L.R. Rand, *Orbital Decay: Space Junk and the Environmental History of Earth's Planetary Borderlands* (Philadelphia, Pennsylvania: University of Pennsylvania, 2016)
29  B. Wilson and D. Kaiser, "Calculating Times: Radar, Ballistic Missiles, and Einstein's Relativity", in *Science and Technology in the Global Cold War*, N. Oreskes N and J. Krige (Eds.) (Boston: MIT Press, 2014).
30  Overhage and Radford, *op. cit*. (Note 22), p. 452.
31  M. Geselowitz, "Paul Rosen: An Interview Conducted by Michael Geselowitz, IEEE History Center, 22 April 2004", https://ethw.org/Oral-History:Paul_Rosen. Rosen (1922-2010) was head of the ground equipment satellite group at Lincoln Laboratory at the time of the PWF tests. Rosen may have confused the 84-foot Millstone Hill Radar antenna with the 60-foot Westford Radio Telescope at Haystack. The latter was an X-band radar antenna purpose-built for use in PWF.
32  W. Liller, "Report on the effects of Project West Ford on optical astronomy", *The Astronomical Journal*, 66 (1961), pp. 114–116.
33  W. Turski, "About the West Ford Project", *Postepy Astronomii Krakow*, 10 (1962), pp. 125–131.
34  A modern estimate of the luminance of the natural night sky is ~200 μcd m$^{-2}$, so the effect would be slightly smaller at about 0.8%. For the luminance figure, see J.C. Barentine, "Night sky brightness measurement, quality assessment and monitoring", *Nature Astronomy*, 6 (2022), p. 1121.
35  Liller (*op. cit.*, Note 32, p. 115) suggested that the force that might align the dipoles with the Earth's magnetic field lines resulted from the net electric charge they would acquire from "the action of the photoelectric effect and collisions with ions, electrons, and other charged particles."
36  Overhage and Radford, *op. cit*. (Note 22), p. 452. Fred Whipple, who later served on the committee established by the Space Science Board to investigate the potential impacts of PWF on astronomy, worked on the metallic chaff project for the then-Army Air Force during the Second World War. Whipple "co-invented a cutter that would turn 3 ounces of aluminum foil into 3000 half-wave dipoles, and he also found optimum aspect ratios for the foil strips that would work over a range of radar frequencies." (D. Brownlee and P. Hodge, "Fred Lawrence Whipple", *Physics Today*, 58 (2005), 88). See also the discussion of this project in "Interview of Fred Whipple by David DeVorkin on 1977 April 29", Niels Bohr Library & Archives, American Institute of Physics, College Park, MD USA, www.aip.org/history-programs/niels-bohr-library/oral-histories/5403.
37  J. Findlay, "West Ford and the scientists", *Proceedings of the IEEE*, 52 (1964), pp. 455–460.
38  Findlay, *op. cit.* (Note 37), p. 455.
39  Findlay, *op. cit.* (Note 37), p. 455.
40  Rand, *op. cit.* (Note 28), p. 87. Rand cites as the source of this quote: "Memo on the April 15th 1960 Meeting of the Space Science Board Ad Hoc Committee on Project NEEDLES," April 15, 1960, 14.B Outer Space 14.B.19 Project Needles 1960; General Records of the Department of State, 1763 - 2002, Record Group 59 Box 347, National Archives at College Park, MD."
41  It is worth noting the definitive detection of the cosmic microwave background by Penzias and Wilson was made only four years later using a 6-meter horn antenna originally built to detect radio waves reflected from the Project Echo satellites. See R.W. Wilson, "Discovery of the cosmic microwave background" in *Modern cosmology in retrospect, sous la dir. de Bruno Bertotti* (Cambridge: Cambridge University Press, 1990), p. 291.
42  G. Hardin, "The tragedy of the commons", *Science*, 162 (1968), pp. 1243–1248.
43  Rand, *op. cit*. (Note 28), pp. 91-92.
44  Findlay, *op. cit.* (Note 37), p. 456.
45  Findlay, *op. cit.* (Note 37), p. 456.
46  Levin, *op. cit*. (Note 23), p. 8.
47  R. Bulkeley, *Origins of the International Geophysical Year* (Berlin: Springer, 2010)
48  Rand (*op. cit.*, Note 28, p. 100, footnote 220) pointed out that the name change resulted from State Department concerns that 'Needles' sounded threatening and might imperil public support for the project: "In a memo to an Operations Coordinating Board official issued mere days after the revised SSB committee report, Raymond Courtney of the State Department detailed the necessity of public support for Project Needles and suggested methods to spread that support. His final recommendation: that the name of the project be changed to something 'less ominous' than 'Needles.' His suggestions include 'Sugar Candy' and 'Cobweb.'" Irwin Shapiro recalled that Edward Purcell suggested 'West Ford', "which was bland enough and stuck" (private communication, 30 December 2023).

49  Findlay, *op. cit.* (Note 37), p. 456.
50  A.E. Lilley, "Long-range radio communication by satellite microwave dipoles: The West Ford project", Nature, 191 (1961), pp. 1237–1238.
51  J. Findlay, *Proceedings of the XIIth URSI General Assembly* (Ghent, Belgium: International Union of Radio Science, 1960), pp. 20–31.
52  Findlay, *op. cit.* (Note 37), p. 457.
53  I.I. Shapiro and H. M. Jones, H. M. "Lifetimes of Orbiting Dipoles", *Science*, 134 (1961), pp. 973–979.
54  Shapiro and Jones, *op. cit.* (Note 53), p. 973.
55  L. Goldberg, "Project West Ford-Properties and Analyses: Introduction", *Astronomical Journal*, 66 (1961), p. 105.
56  Lilley, *op. cit.* (Note 50), p. 1237.
57  Liller, *op. cit.* (Note 32), p. 114.
58  Liller, *op. cit.* (Note 32), p. 115.
59  Findlay, *op. cit.* (Note 37), p. 457.
60  American Astronomical Society, "Resolution Adopted by American Astronomical Society, June 20, 1961, at Nantucket, Massachusetts", *The Astronomical Journal*, 66 (1961), p. 275.
61  A. McClure, "The vanishing sky", *The Review of Popular Astronomy*, 55 (1961), pp. 8–11.
62  McClure, *op. cit.* (Note 61), p. 8.
63  McClure, *op. cit.* (Note 61), p. 11.
64  S. Mesics, "Satellite pollution", *Lehigh Valley Amateur Astronomical Society Observer*, 60 (2020), p. 9. Johnson's letter to McClure is reproduced there in whole.
65  H. Bondi, "West Ford Project, Introductory Note by the Secretary", *Quarterly Journal of the Royal Astronomical Society*, 3 (1962), p. 99.
66  D. Alpert, L. Goldberg, K.M. Watson, G.L. Salton and D.H. Cooper, West Ford Communication Techniques: Cost, Effectiveness and Potential vs. Competing Systems. Technical report ADA955434 (Washington. D.C.: Office of the Director of Defense Research and Engineering, 1965), p. 16. https://apps.dtic.mil/sti/tr/pdf/ADA955434.pdf
67  Findlay, *op. cit.* (Note 37), p. 457.
68  Lilley, *op. cit.* (Note 50), p. 1238.
69  Findlay, *op. cit.* (Note 37), p. 458.
70  J.C. Pecker and G. Ringeard, "Astronomy, the People and the Governments", *Vistas in Astronomy*, 15 (1973), pp. 1–12.
71  J. Denisse, "40. Commission de Radio-Astronomie. Compte rendu des Séances", *Transactions of the International Astronomical Union*, 11 (1962), pp. 349–350.  The named individuals were all astronomers and members of IAU Commission 40.
72  Denisse, *op. cit.* (Note 71), p. 349.
73  Denisse, *op. cit.* (Note 71), p. 355.
74  Denisse, *op. cit.* (Note 71), p. 354.
75  Alpert, Goldberg, Watson, Salton and Cooper, *op. cit.* (Note 66), p. 20.
76  Findlay, *op. cit.* (Note 37), p. 458.
77  J. Turner, "Two cheers for West Ford", *Science*, 134 (1961), p. 641.
78  Sunlight falling on objects with a relatively large area-to-mass ratio like the PWF dipoles yields radiation pressure that exerts a force that modifies their orbits; "for certain ('resonant') combinations of orbital altitudes and inclinations the effects of solar radiation essentially build up monotonically, seriously affecting the [orbital] lifetime." (R.W. Parkinson, H. M. Jones and I.I. Shapiro, "Effects of Solar Radiation Pressure on Earth Satellite Orbits", *Science*, 131 (1960), p. 920). The parameters of the PWF dipoles' orbit were chosen to exploit this effect in order to deliberately limit their time in orbit and hasten their re-entry into the Earth's atmosphere.
79  Findlay, *op. cit.* (Note 37), p. 457.
80  Findlay, *op. cit.* (Note 37), p. 457.

81 NSSDCA/COSPAR ID 1961-028A.
82 Overhage and Radford, *op. cit*. (Note 22), p. 453.
83 National Aeronautics and Space Administration, MIDAS 4. Technical Report Version 5.1.15. (2022) https://nssdc.gsfc.nasa.gov/nmc/spacecraft/display.action?id=1961-028A.
84 D.S. Greenberg, "Project West Ford: Cause of failure still unknown", *Science*, 134 (1961), p. 1680.
85 C. Wiedemann, J. Bendisch, H. Krag, P. Wegener and D. Rex, "Modeling of copper needle clusters from the West Ford Dipole experiments", In *Proceedings of the Third European Conference on Space Debris, 19-21 March 2001, Darmstadt, Germany*, ESA SP-473, Vol. 1 (Noordwijk, Netherlands: ESA Publications Division, 2001).
86 Greenberg, *op. cit*. (Note 84), p. 1680.
87 MIT Lincoln Laboratory, "Status Report on Project West Ford, March 1, 1962", *URSI Information Bulletin*, 130 (1962), p. 5012; Findlay, op. cit. (Note 37), p. 458.
88 Findlay, op. cit. (Note 37), p. 459.
89 Rand, op. cit. (Note 28), p. 113.
90 United States of America, "Project West Ford Experiment", *Position Paper Annex G. Technical report, U.N. Committee on the Peaceful Uses of Outer Space* (1962), p. 2. https://www.cia.gov/readingroom/docs/CIA-RDP66R00638R000100150083-6.pdf.
91 A.C.B. Lovell and M. Ryle, "Interference to Radio Astronomy from Belts of Orbiting Dipoles (Needles)", *Quarterly Journal of the Royal Astronomical Society*, 3 (1962), pp. 100-108.
92 D.E. Blackwell and R. Wilson, "Interference to Optical Astronomy from Belts of Orbiting Dipoles (Needles)", *Quarterly Journal of the Royal Astronomical Society*, 3 (1962), pp. 109–114.
93 Bondi, *op. cit*. (Note 65), p. 99.
94 E. Purcell, "The case for the 'needles' experiment", *New Scientist,* 13 (1962), pp. 245–247.
95 NSSDCA/COSPAR ID 1962-Kappa-2.
96 I.I. Shapiro, private communication, 30 December 2023.
97 R. Meade, *Operation Fishbowl. Technical report LA-UR-22- 31336* (Washington, D.C.: U.S. Department of Energy National Nuclear Security Administration, 2022), p. 1. https://doi.org/10.2172/1896391
98 F. Narin, *A "Quick Look" at the Technical Results of Starfish Prime* (Washington, D.C.: Defense Technical Information Center, 1962), pp. 3-4. https://apps.dtic.mil/sti/pdfs/ADA955411.pdf
99 "[A]t least eight satellites suffered damage that was definitely related to the STARFISH PRIME event." E.E. Conrad, G.A. Gurtman, G. Kweder, M.J. Mandell and W.W. White (2010) *Collateral damage to satellites from an EMP attack, Technical report ADA531197* (Washington, D.C.: Defense Threat Reduction Agency, 2010), pp. 11-15. https://apps.dtic.mil/sti/pdfs/ADA531197.pdf
100 "The grave results of this test meant that the conditions that initially suggested using a field of passive reflectors for a nuclear-proof military communications backup in 1962 had become even more serious, and justified the need for a second West Ford test. The Air Force moved forward with plans for a second launch in late 1962." Rand, *op. cit.* (Note 28), pp. 114-115.
101 Rand, *op. cit.* (Note 28), p. 117.
102 NSSDCA/COSPAR ID 1963-014A.
103 National Aeronautics and Space Administration, MIDAS 6. Technical Report Version 5.1.15 (2022). https://nssdc.gsfc.nasa.gov/nmc/spacecraft/display.action?id=1963-014A.
104 R.L. Smith-Rose, *URSI Information Bulletin*, 147 (1962), pp. 41–42.
105 Overhage and Radford, *op. cit*. (Note 22), p. 454.
106 Findlay, *op. cit.* (Note 37), p. 459.
107 "Lincoln Laboratory provided look angles and transmissions, but for some reasons still unknown the scattered signals could not be detected, although the predicted levels should have been several times greater than the radiometer noise. There were experimental problems at the receiving end in maintaining accurate frequency control and in ensuring that the two radio beams overlapped on the belt. The NRAO telescope was unable to track or search for the illuminated patch of the belt, so that observations were attempted by leaving the NRAO instrument at a predetermined pointing position while Millstone tracked the belt in such a way that the illuminated patch should have passed through the beam of the receiving antenna." Findlay, *op. cit.* (Note 37), p. 460.
108 Figure 1 from *Project West Ford: A feasibility study of orbital scatter communication, by the Massachusetts Institute of Technology, Lincoln Laboratory, Lexington, Massachusetts, March 27, 1964. Satellite Communications, 1964: Hearings Before a Subcommittee of the Committee on Government*

*Operations, House of Representatives, Eighty-eighth Congress, Second Session, Volume 74-77* (Washington, D.C.: U.S. Government Printing Office, 1964), p. 542.

109 I.I. Shapiro II and H.M. Jones, "Lifetimes of orbiting dipoles", *Science*, 134 (1961), pp. 973–979.

110 W.G. Tifft, W.M. Sinton, J.B. Priser and A.A. Hoag, "West Ford dipole belt: Photometric observations", *Science*, 141 (1963), pp. 798–799.

111 I.I. Shapiro, H.M. Jones and C. Perkins, "Orbital properties of the West Ford dipole belt", *Proceedings of the IEEE,* 52 (1964), pp. 469– 518.

112 A. Sandage and C. Kowal, "West Ford dipole belt: Optical detection at Palomar", *Science*, 141 (1963), pp. 797–798.

113 Findlay, *op. cit.* (Note 37), p. 460.

114 Sandage and Kowal, *op. cit.* (Note 112), p. 798.

115 *New York Times*, 16 June 1963, 2.

116 Findlay, *op. cit.* (Note 37), p. 460.

117 American Physical Society, "Project West Ford", *Physics Today*, 17 (1964), p. 76.

118 Replotted from Figure 1 in W. Liller, "Optical effects of the 1963 project West Ford experiment", *Science*, 143 (1964), p. 439.

119 Barentine, *op. cit.* (Note 34), p. 1121.

120 J.W.E. Morrow and D.C. MacLellan, "Properties of orbiting dipole belts", *The Astronomical Journal,* 66 (1961), p. 107.

121 Blackwell and Wilson, *op. cit.* (Note 92), pp. 113-114.

122 Findlay, *op. cit.* (Note 37), p. 460.

123 Committee on Space Research, "Project West Ford", *COSPAR Information Bulletin*, 20 (1964), pp. 25–26.

124 Alpert, Goldberg, Watson, Salton and Cooper, *op. cit.* (Note 66), p. 13.

125 Not all of the dipoles re-entered. Clumps of needles from both PWF test launches remain in orbit and can be tracked from the ground using radar. See, e.g., R.M. Goldstein, S.J. Goldstein, and D.J. Kessler, "Radar observations of space debris", *Planetary and Space Science*, 46 (1998), p. 1013: "Even though some authors have maintained that all of the needles have reentered, new lumps of needles have been catalogued years after he West Ford Project."

126 I.I. Shapiro, "Last of the West Ford dipoles", *Science*, 154 (1966), pp. 1445–1448.

127 Rand noted that "Chuck Perkins, who worked on the West Ford design as a graduate student at MIT, later gathered samples from a region of Antarctica where the needles were likely to have fallen. These samples showed no evidence of dipole detritus." (*op. cit.*, Note 28, p. 89, footnote 200).

128 Shapiro, *op. cit.* (Note 126), p. 1448.

129 J.M. Kemp, *Evolution toward a space treaty: an historical analysis, Technical report HHN*-64 (Washington, D.C: NASA Office of Policy Analysis, 1966), p. 6. https://ntrs.nasa.gov/api/citations/19680006069/downloads/19680006069.pdf

130 United States of America, *op. cit.* (Note 90), pp. 1-2.

131 A. Kerrest, "Outer Space as International Space: Lessons from Antarctica", in *Science Diplomacy: Antarctica, Science, and the Governance of International Spaces* (P.A. Berkman, M.A. Lang, D.W.H. Walton and O.R. Young, Eds.) (Washington, D.C.: Smithsonian Institution Scholarly Press, 2011), p. 139.

132 R.A. Musto, "Antarctic arms control as past precedent", *Polar Record*, 55 (2019), pp. 330–333.

133 C. Collis, "Territories beyond possession? Antarctica and outer space", *The Polar Journal*, 7 (2017), pp. 287–302.

134 D.D. Eisenhower, *Public Papers of the President of the United States: Dwight D. Eisenhower, 1960–61* (Washington, D.C.: U.S. Government Printing Office, 1962), p. 714.

135 United States Department of State, *Department of State Publication 7413* (Washington, D.C.: U.S. Government Printing Office, 1962), p. 87.

136 United Nations General Assembly Resolution 1884 (XVIII), 17 October 1963.

137 United Nations General Assembly Resolution 1962 (XVIII), 13 December 1963; see B. Gupta B and E. Rathore, "United nations general assembly resolutions in the formation of the outer space treaty of 1967", *Astropolitics*, 17 (2019), pp. 77–88.

138 P.G. Dembling and D.M. Arons, "The evolution of the outer space treaty", *Journal of Air Law and Commerce*, 33 (1967), pp. 419–456.

139 "Treaty on Principles Governing the Activities of States in the Exploration and Use of Outer Space, including the Moon and Other Celestial Bodies", 18 U.S.T. 2410 610 U.N.T.S. 205, 61 I.L.M. 386 (1967).
140 A.C.B. Lovell , "The Effects of Defence Science on the Advance of Astronomy", *Journal for the History of Astronomy*, 8 (1977), p. 151.
141 Figure 1 in Pecker and Ringeard, *op. cit.* (Note 70), p. 5.
142 Rand, *op. cit.* (Note 28), p. 95.
143 Rand, *op. cit.* (Note 28), p. 134.
144 M. Byers and A. Boley, *Who Owns Outer Space? International Law, Astrophysics, and the Sustainable Development of Space* (Cambridge: Cambridge University Press, 2023), p. 104.
145 McDonald and Naugle, *op. cit.* (Note 10), p. 363.
146 Massey, Lucatello and Benvenuti, *op. cit.* (Note 1), p. 1022. In setting up a discussion about "the natural tension between scientific and commercial use" of the terrestrial radio spectrum, the authors note that "as early as 1961, radio astronomers warned of the threat the exploitation of space might pose to our science when Project West Ford attempted to create an artificial ionosphere to enhance communications by deploying 480 million copper needles in orbit."
147 O.R. Hainaut and A.P. Williams, "Impact of satellite constellations on astronomical observations with ESO telescopes in the visible and infrared domains", *Astronomy & Astrophysics*, 636 (2020), p. 1; R. Ragazzoni, "The Surface Brightness of MegaConstellation Satellite Trails on Large Telescopes", *Publications of the Astronomical Society of the Pacific*, 132 (2020), p. 114502.
148 See, e.g., S. Kruk, P. García-Martín, M. Popescu, B. Aussel, S. Dillmann, M.E. Perks, T. Lund, B. Merín, R. Thomson, S. Karadag and M.J. McCaughrean, "The impact of satellite trails on Hubble Space Telescope observations", *Nature Astronomy*, 7 (2023), pp. 262–268.
149 M. Kocifaj, F. Kundracik, J.C. Barentine and S. Bará, "The proliferation of space objects is a rapidly increasing source of artificial night sky brightness", *Monthly Notices of the Royal Astronomical Society: Letters*, 504 (2021), pp. L40–L44; C.G. Bassa, O.R Hainaut, and D. Galadí-Enríquez, "Analytical simulations of the effect of satellite constellations on optical and near-infrared observations", *Astronomy & Astrophysics*, 657 (2022), p. A75.
150 J.A. Tyson, Ž. Ivezić, A. Bradshaw, M.L. Rawls, B. Xin, P. Yoachim, J. Parejko, J. Greene, M. Sholl, T.M.C. Abbott, and D. Polin, "Mitigation of LEO Satellite Brightness and Trail Effects on the Rubin Observatory LSST", *The Astronomical Journal*, 160 (2020), p. 226; J.C. Barentine, A. Venkatesan, J. Heim, J. Lowenthal, M. Kocifaj and S. Bará, "Aggregate effects of proliferating low-Earth-orbit objects and implications for astronomical data lost in the noise", *Nature Astronomy*, 7 (2023), pp. 252–258.
151 H. Xie and H. Minn, "Analysis of RFI from Mega-Constellation NGSO Communication Satellites to Ground Radio Astronomy Systems", *IEEE Transactions on Communications*, in press (2024); F. Di Vruno, B. Winkel, C.G. Bassa, G.I.G. Józsa, M.A. Brentjens, A. Jessner, and S. Garrington, "Unintended electromagnetic radiation from Starlink satellites detected with LOFAR between 110 and 188 Mhz", *Astronomy & Astrophysics*, 676 (2023), p. A75; and C.G. Bassa, F. Di Vruno, B. Winkel, G.I.G. Józsa, M.A. Brentjens and X. Zhang, "Bright unintended electromagnetic radiation from second-generation Starlink satellites", *Astronomy & Astrophysics*, 689 (2024), p. L10.
152 S. Solter-Hunt, "Potential perturbation of the ionosphere by megaconstellations and corresponding artificial re-entry plasma dust", arXiv (2023), https://arxiv.org/abs/2312.09329.
153 D.M. Murphy, M. Abou-Ghanem, D.J. Cziczo, K.D. Froyd, J. Jacquot, M.J. Lawler, C. Maloney, J.M.C. Plane, M.N. Ross, G.P. Schill, and X. Shen, X. "Metals from spacecraft reentry in stratospheric aerosol particles", *Proceedings of the National Academy of Sciences*, 120 (2023), e2313374120.
154 A.C. Boley and M. Byers, "Satellite mega-constellations create risks in Low Earth Orbit, the atmosphere and on Earth", *Scientific Reports*, 11 (2021), p. 10642.
155 C. Walker and J. Hall (Eds.), "Impact of Satellite Constellations on Optical Astronomy and Recommendations Toward Mitigations", *Bulletin of the American Astronomical Society*, 52 (2021), no. 0206; C. Walker (Ed.), *Dark and Quiet Skies for Science and Society On-line Workshop: Report and recommendations*, https://doi.org/10.5281/zenodo.5898785.
156 G.I.G. Józsa, G. I. G., *et al.* "Call to Protect the Dark and Quiet Sky from Harmful Interference by Satellite Constellations", arXiv (2024), https://doi.org/10.48550/ARXIV.2412.08244.
157 G. Rotola and A. Williams, "Regulatory Context of Conflicting Uses of Outer Space: Astronomy and Satellite Constellations", *Air and Space Law*, Vol. 46 (2021), pp. 545–568; D.A. Koplow, "Blinded by the Light: Resolving the Conflict between Satellite Megaconstellations and Astronomy", *Vanderbilt Journal of*

*Transnational Law*, 53 (2023), pp. 219-299; and S. Freeland and A.-S. Martin, "A Sky Full of Stars, Constellations, Satellites and More! Legal Issues for a 'Dark' Sky", *Oslo Law Review*, 10 (2024), pp. 1–22).

158 Chile, Spain, Slovakia, Bulgaria, Dominican Republic, Peru, South Africa, International Astronomical Union, European Organization for Astronomical Research in the Southern Hemisphere and Square Kilometre Array Observatory, "Conference Room Paper on the Protection of Dark and Quiet Skies for science and society", Committee on the Peaceful Uses of Outer Space Scientific and Technical Subcommittee A/AC.105/C.1/2023/CRP.18/Rev.1 (2023); Argentina, Austria, Belgium, Bulgaria, Chile, Colombia, Czechia, Denmark, Ecuador, Germany, Italy, Netherlands, Paraguay, Peru, Slovakia, South Africa, Spain, Switzerland, European Astronomical Society, European Astronomical Research in the Southern Hemisphere, International Astronomical Union and Square Kilometre Array Observatory, "Conference Room Paper on the Protection of Dark and Quiet Skies for science and society", Committee on the Peaceful Uses of Outer Space Scientific and Technical Subcommittee A/AC.105/C.1/2024/CRP.18 (2024).

159 See, e.g., S.E. Grotch, "Mega-Constellations: Disrupting the Space Legal Order", *Emory International Law Review*, 37 (2022), pp. 102–134; and S. Millwood, *The Urgent Need for Regulation of Satellite Mega-constellations in Outer Space*. (Cham, Switzerland: Springer International Publishing, 2023).

160 A. Venkatesan, J. Lowenthal, P. Prem, and M. Vidaurri, "The impact of satellite constellations on space as an ancestral global commons", *Nature Astronomy*, 4 (2020), pp. 1044-1045.

161 Venkatesan, Lowenthal, Prem and Vidaurri, *op. cit.* (Note 160), p. 1045.